\documentclass[journal]{IEEEtran}

\usepackage{amsmath,amsfonts}
\usepackage{algorithmic}
\usepackage{algorithm}
\usepackage{array}
\usepackage[caption=false,font=normalsize,labelfont=sf,textfont=sf]{subfig}
\usepackage{textcomp}
\usepackage{stfloats}
\usepackage{url}
\usepackage{verbatim}
\usepackage{graphicx}
\usepackage{cite}
\usepackage[T1]{fontenc}
\usepackage{color, soul}
\usepackage{comment}
\usepackage{amssymb}
\usepackage{textgreek}
\usepackage{tcolorbox}
\usepackage{enumitem}
\usepackage{tabularx}
\usepackage{makecell}
\usepackage{multirow}
\usepackage{flushend}
\usepackage{xurl}
\usepackage[hidelinks]{hyperref}
\usepackage{adjustbox}

\usepackage{booktabs}
\usepackage{ragged2e}

\usepackage{balance}

\usepackage{hyperref}
\hypersetup{
    pdftitle={From Grid to Chip: Power Architecture, Stability, and Flexibility of AI Data Centers},   
    pdfauthor={},    
    pdfsubject={},   
    pdfcreator={},   
    pdfproducer={},  
    pdfkeywords={},  
}

\newcolumntype{C}[1]{%
    >{\hsize=#1\hsize
      \linewidth=\hsize
      \Centering\arraybackslash}X%
}
\newcommand{\doubletoprule}{%
  \toprule
  \specialrule{\lightrulewidth}{\doublerulesep}{\doublerulesep}
}
\newcommand{\doublebottomrule}{%
  \specialrule{\lightrulewidth}{\doublerulesep}{\doublerulesep}
  \bottomrule
}



\begin{document}

\bstctlcite{IEEEexample:BSTcontrol} 

\title{From Grid to Chip: Power Architecture, Stability,\\ and Flexibility of AI Data Centers}

\author{Yubo~Song,~\textit{Senior~Member,~IEEE}, Rui~Kong,~\textit{Member,~IEEE}, Takuro~Umihara,~\textit{Student~Member,~IEEE}, Pooya~Davari,~\textit{Senior~Member,~IEEE}, Frede~Blaabjerg,~\textit{Fellow,~IEEE}, Subham~Sahoo,~\textit{Senior~Member,~IEEE}

\thanks{This work is supported in part by Aalborg University (AAU) through the initiative on Power2Compute (P2C), in part by the European Research Council (ERC) Synergy Grants through the project Store-And-Forward Energy Grid (SAFEr Grid), and in part by the Innovation Fund Denmark (IFD) through the project Artificial Intelligence for Next Generation Power Electronics (AI-Power).}

\thanks{All authors are with the Department of Energy, Aalborg University, Aalborg, Denmark. (e-mails: \{\texttt{yuboso, ruko, taum, pda, fbl, sssa}\}@energy.aau.dk) (\textit{Corresponding author: Subham Sahoo})}
}

\markboth{IEEE Open Journal of Power Electronics}%
{Shell \MakeLowercase{\textit{et al.}}: A Sample Article Using IEEEtran.cls for IEEE Journals}


\maketitle

\begin{abstract}
The rapid growth of artificial intelligence (AI) computing is transforming data centers into large, dynamic electrical loads. Their deployment is primarily constrained by energy availability and grid-connection capacity, which is further aggravated by the ability of power-delivery architectures, control systems, and computing workloads to operate reliably during fast grid disturbances. This article provides a comprehensive overview of forefront challenges and trends in the grid-to-chip framework, and presents a technological perspective on AI data centers as \textit{grid-interactive computing systems}. First, it reviews grid-integration bottlenecks, evolving connection policies, grid-code requirements, which has fostered new technological trends via spatio-temporal flexibility available through workload orchestration, cooling systems, on-site resources, and energy storage. Second, it maps the evolution of power-delivery architectures from medium-voltage grid interfaces to chip-level, discussing higher-voltage DC distribution, solid-state transformers, wide-bandgap devices, advanced chip-level power delivery, and liquid cooling. Third, it establishes a three-level stability framework spanning rack-level DC-bus dynamics, facility-level converter interactions, and system-level grid-coupled behavior. The framework connects dominant instability mechanisms, including constant power load effects, impedance interactions, forced oscillations, and operating-mode transitions, with suitable modeling, assessment, and mitigation approaches. Synthesizing these topics, this article highlights grid-to-chip co-design as a central requirement for scalable AI infrastructure, linking computing workloads, power-delivery systems, energy buffers, and grid operation. 
\end{abstract}


\section{Introduction}

\IEEEPARstart{T}{he} rapid growth of artificial intelligence (AI) is making data centers a major and increasingly concentrated source of electricity demand. The computational resources used to train AI models have expanded rapidly, while global data center electricity consumption is projected to increase substantially during this decade \cite{OpenAI,IEA_KeyQuestions}. This expansion is increasingly constrained by limited network capacity, long interconnection lead times, shortages of grid equipment delivery, and local infrastructure constraints \cite{JLL_2026,IEA_pause,IEA_trans}. The resulting challenge is therefore not simply to supply more electrical energy, but also to integrate high-density AI computing infrastructure into power systems reliably, flexibly, and sustainably.

AI data centers differ from conventional large electrical loads in both power density and dynamic behavior. Accelerators-dominated computing can produce rapid and correlated changes in power demand, whereas the response observed by the grid is shaped by cascaded power electronic converters, energy-storage systems, cooling equipment, protection logic, and internal distribution networks \cite{nvidia_smoothing,Choukse2025PowerStabilization,ornl2026,ocp2026lvdc}. Consequently, annual energy demand and peak-load estimates alone are insufficient to characterize their system impact. The electrical behavior of an AI data center must be considered across the full path from chip-level voltage regulation to the point of common coupling.

This calls for a systemic reliability evaluation that must be assessed at the level of the coupled system rather than from the availability of individual servers, uninterruptible power supplies (UPS), or grid assets alone. The relevant dynamics span controlled power converters, constant power-approximated IT loads, energy buffers, protection and transfer logic, internal distribution networks, and the external grid. Reported events have demonstrated that converter and load interactions can produce facility- or grid-level oscillations, while disturbance-driven load disconnection and restoration can cause substantial frequency and voltage excursions \cite{NERC2026RiskMitigation,EirGridSONI2025LargeDemandFRT}. These observations motivate stability assessment across multiple physical boundaries and time scales.

Although prior work has examined AI-related electricity demand, large-load interconnection, flexible computing, data-center power architectures, and converter-dominated stability, these topics are often addressed at separate physical and institutional boundaries. A unified perspective is needed because workload decisions, power-delivery architecture, energy buffering, protection behavior, and grid conditions jointly determine both the electrical behavior and the scalable deployment of AI infrastructure. Hence, this article provides a comprehensive overview of forefront challenges and trends in the grid-to-chip framework, extensively synthesizing power-system integration, power-electronics architecture, workload flexibility, and dynamic stability. Critically, this article offers insights in treating AI data centers as \textit{grid-interactive computing systems}, and highlights grid-to-chip co-design that aligns future scalable AI infrastructure with grid advancements.

Specifically, Section II reviews the bottlenecks, connection arrangements, grid-code developments, and flexibility mechanisms relevant to data-center integration. Section III maps the evolution of power-delivery architectures from the grid interface to chip-level regulation. Section IV develops a three-level stability framework covering rack-level, facility-level, and system-level interactions, together with modeling, assessment, and enhancement approaches. Section V outlines future directions for grid-interactive computing infrastructure, evolving grid architectures, convergent power-information systems, and sustainable data-center operation.

\section{Integration of Data Centers into Power Grids: Primary Challenges}

With rapid advances in the performance of AI models and the growing number of users, a large number of data center construction has been planned. At the same time, many projects have been suspended due to bottlenecks in power system operation and supply chain integration. This chapter describes the current status of the integration of data centers into power grids and discusses the challenges and opportunities associated with this integration.

\subsection{Today's Data Center Integration Status}

\subsubsection{Surging Data Center Demand}

\begin{figure}[t]
    \centering
    \includegraphics[width=\linewidth]{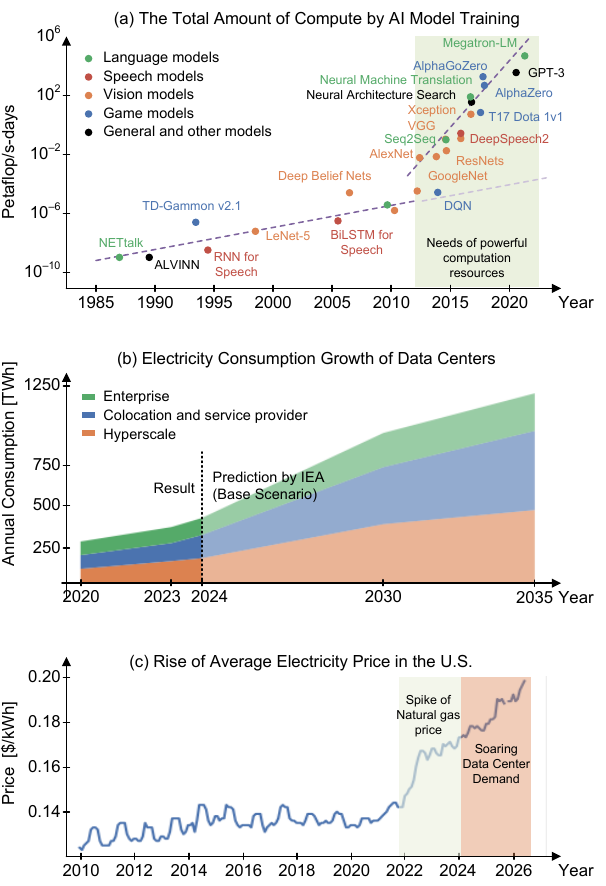}
    \caption{Demand increase trend of data centers. (a) Computational resources used by each AI model training \cite{OpenAI}, (b) Global electricity demand growth of data centers \cite{IEA_AI}, (c) Rise of average electricity price in the U.S. \cite{FRED}.} \label{fig:demand}
\end{figure}

With advances in AI models, the need to integrate data centers into power systems is rapidly increasing. Fig.~\ref{fig:demand} shows the trends in the amount of computing resources used for AI model training, the trends in data center electricity consumption and future projections by the IEA as a base-case scenario, and the trends in average residential electricity rates in the United States. With the emergence of image recognition models based on convolutional neural networks (CNN) around 2012 and the advent of large language models (LLMs) around 2019, the amount of computational resources used for AI training has increased exponentially, with a doubling time of 3.4 months since 2012 \cite{OpenAI}. In addition, LLM services are rapidly gaining popularity. For ChatGPT, the average number of daily user messages was 451 million in June 2024, rising to 2627 million by June 2025 \cite{Aaron_2025}. Along with this rapid increase in computational demand and the growing popularity of these services, data center electricity consumption is also expanding, as shown in Fig.~\ref{fig:demand} (b). The annual electricity consumption of operational data centers stood at 416~TWh as of 2024. The breakdown is as follows: 106~TWh for enterprise use, 144~TWh for colocation and service providers, and 166~TWh for hyperscale operations. The IEA projects that under its base-case scenario, total consumption will reach 946~TWh by 2030, and 1193~TWh by 2035 \cite{IEA_AI}.

Connecting large-scale data centers has increased electricity demand and led to supply shortages, which in turn could drive up electricity price rates, as shown in Fig.~\ref{fig:demand} (c). In fact, average residential electricity rates in the US have continued to rise since 2022. Inflation from 2022 to 2024 was mainly driven by sharp spikes in natural gas prices following Russia's invasion of Ukraine. The rise in rates since 2024 was partly attributed to increased data center connection capacity. Analysis in \cite{FRBD_DC} pointed out that operational data centers have already increased electricity wholesale prices by 2\% to 6\% on average nationwide in the US between 2021 and 2025, with larger effects in regions hosting major data center corridors.

\subsubsection{Bottlenecks of Data Center Integration}

While many new data center projects have been announced, construction delays have been reported around the world. 
Table~\ref{tab:leadtime} compares data center capacity (both operational and planned) in major data center corridors\cite{Kushman_2026}, the power synchronous systems to which they belong, the lead time from planning to integration \cite{JLL_2026}, and construction costs \cite{JLL_2026} in each area. 
Amsterdam and Tokyo have the longest lead times, at 10 years, for new data center projects. Moreover, Amsterdam has paused new data center projects due to insufficient grid capacity to integrate large new power loads \cite{IEA_pause}.

\begin{table*}[t]
    \centering
    \renewcommand{\arraystretch}{1.2}
    \caption{Comparison of Industrial Regions on Data Center Development Capacity, Lead Time,\\and Construction Costs for Data Center Integration}
    \label{tab:leadtime}
    \vspace{-6pt}
    \begin{tabular}{c|c|c|c|c|c}
    \hline \hline
     Region & Synchronous & Operational Cap.  & Planned Cap. & Lead Time & Construction Costs \\
      & Grid Area & [MW]\cite{Kushman_2026} &  [MW]\cite{Kushman_2026} &  [years] \cite{JLL_2026}&  [\$ Millions/MW] \cite{JLL_2026}\\
    \hline 
    Amsterdam, Netherlands & RG Continental Europe & 850 & 250 & 10 & 11-12 \\ \hline
    Tokyo, Japan & Eastern Japan & 1200 & 1705 & 10 & 14-18 \\ \hline
    Frankfurt, Germany & RG Continental Europe & 800 & 1228 & 7 & 11-13 \\ \hline
    London, UK & RG Great Britain &1290 & 1300 & 7 & 12-14 \\ \hline
    Virginia, US & Eastern Interconnection & 11300 & 37500 & 7 & 11-12 \\ \hline
    Atlanta, US & Eastern Interconnection & 1800 & 10000 & 5 & 10-11 \\ \hline
    Singapore & Singaporean Grid & 1000 & 237 &  4 & 12-15 \\ \hline
    Sydney, Australia & National Electricity Market & 750 & 1102 & 3 & 11-12 \\ \hline
    Dallas, US & Texas Interconnection & 1800 & 10000 & 2.5 & 10-11 \\ \hline
    Mumbai, India & Indian Grid & 750 & 998 & 2 & 6-7 \\ \hline
    Dublin, Ireland & RG Ireland & 1260 & 695 &  - & 10-11 \\ \hline \hline
    \multicolumn{6}{l}{"-": Not specified due to pausing new construction projects.} \\
    \end{tabular}
\end{table*}

\begin{figure}[t]
    \centering
    \includegraphics[width=1\linewidth]{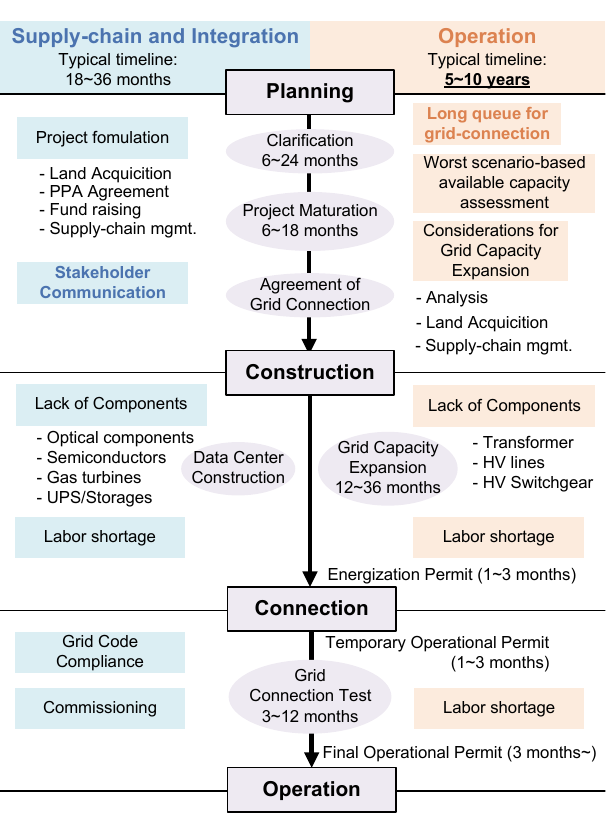}
    \caption{Bottleneck analysis of data center integration. PPA: Power Purchase Agreement}
    \label{fig:bottleneck}
\end{figure}

The reasons for the long lead time include bottlenecks that exist throughout the process, from data center planning to the start of operations. Fig.~\ref{fig:bottleneck} illustrates the bottlenecks at each stage. From top to bottom, the figure shows the stages of data center integration: planning, construction, grid connection, and start of operations. The left side of the figure shows bottlenecks in supply chain and integration, while the right side shows bottlenecks in power grid operations. A key point is that while the typical timeline for data center construction is 18 to 36 months, the timeline for power grid operations is 5 to 10 years \cite{FTI_Leadtime, DDCI_leadtime}. In other words, many of the current bottlenecks exist on the operations side.

During the planning phase, the data center developer formulates the project. This includes securing financing and land, as well as managing the supply chain for power purchase agreements (PPAs) and materials. Delays in these tasks can create bottlenecks. In particular, securing electricity and water is critical for the stable operation of a data center.

At the same time, stakeholder communication aimed at reaching agreement on the construction plan is also crucial. In particular, there have been cases where projects were abandoned due to insufficient understanding of data center construction, leading to friction with local residents and other stakeholders. In 2026, the Detroit City Council in Michigan, the US, approved a resolution opposing the unrestricted use of water resources by data centers \cite{Detroit}, and a nearby utility authority suspended water supply to a specific data center for 12 months \cite{YCUA}.

After receiving notification from data center developers regarding grid interconnection plans, power grid system operators clarify the plans, facilitate the project maturation, and reach an agreement on the interconnection plan. The timeframes listed in each step are based on guidelines provided by Energinet, Denmark’s transmission system operator (TSO) \cite{Energinet_Flow}. The same applies to Fig. \ref{fig:bottleneck}. The long queue for this procedure creates a bottleneck that extends lead times. In this process, grid interconnection management has traditionally been operated under a "first-come, first-served" principle, whereby the operator who receives approval first gains priority for grid interconnection. Under this rule, operators rushed to submit applications early in order to secure priority \cite{EU_FCFS}. Several system operators, including those in regions such as Frankfurt and Virginia, as shown in Table~\ref{tab:leadtime}, have revised the first-come, first-served rule and established a "first-ready, first-served" priority system that takes into account the feasibility of construction plans \cite{FERC_PJM}.

The available interconnection capacity calculated by system operators is generally derived from the difference between the allowable thermal capacity of transmission lines and substations and the allocated capacity. Furthermore, the feasibility of connection is evaluated by comparing the available capacity with the "maximum output during the annual peak demand". These calculations do not take into account demand-side flexibility. Consequently, there are cases where new demand that could be connected to the grid by utilizing flexibility is denied grid connection permission on the grounds of insufficient available capacity. An analysis of the U.S. power system \cite{Norris_2025} concludes that by factoring in flexibility capable of curtailing 0.25\% of annual peak demand, 76~GW of new load, equivalent to 10\% of the nation’s current aggregate peak demand, could be integrated into the grid.
If grid expansion were to be considered for data center integration, it would require additional time to secure land and materials and to conduct analyses of economic and environmental impacts.

During the construction phase, shortages of materials and human resources can act as bottlenecks for both data center developers and power grid operators.
Building power infrastructure, such as transformers and power lines, takes longer than constructing a data center. According to the IEA, average lead times for power cables and large transformers have approximately doubled from 2021 to 2024 due to the heavy electrification of society \cite{IEA_trans}. As a result, there are cases in which a data center is built but cannot be connected to the power grid, leading to longer lead times.

During the grid connection phase, both parties will conduct grid connection tests. The data center operator must comply with the grid code established by the system operator. Any unforeseen events that occur during commissioning or grid connection testing may cause delays.

\subsection{Power Systems Challenges and Opportunities on Data Center Integration.}

\subsubsection{Grid Code Regulations}

In preparation for data center consolidation, TSOs and regulatory agencies in each region have been revising grid codes and related regulations. However, the status of these revisions and the specific items being revised vary by TSO globally.
Table~\ref{tab:gridcodes} compares grid codes and related regulations for large-capacity load connections for TSOs in major data center corridors. For comparison, 9 representative TSOs are selected. PJM, ERCOT, and AESO are TSOs in North America. Energinet is the Danish TSO. TenneT operates Dutch and Northern German transmission systems. EirGrid is the Irish TSO. RTE operates transmission systems in France. NESO is the TSO in the UK. TEPCO is the system operator in the Greater Tokyo region in Japan.

The conventional "first-come, first-served" approach to connection queue management is not well-suited to handling a large number of concurrent grid connection requests, some of which are for speculative purposes. First-ready, first-served queue management policy considers the readiness of grid connections for each application, prioritizing those with high feasibility. This policy has been introduced in the UK since 2025 \cite{UK_GridCode}, Germany, and Texas, US, in 2026 \cite{German_GridCode, ERCOT_GridCode}.
The Netherlands, on the other hand, has introduced a social-importance-based grid-connection prioritization rule since January 2026. The new framework prioritizes major social objectives, including residential construction, hospitals, schools, and national defense \cite{Dutch_GridCode}.

Penalties for delayed projects and screening of multiple applications are considered as compliance measures to mitigate infeasible applications. In Japan, a penalty applies to demand over 30~MW; if a customer continues for a certain period without setting the final contracted capacity as originally planned, the general transmission and distribution operator will review the contracted capacity and impose a penalty \cite{Japan_Penalty}. Screening has been widely introduced by TSOs. It aims to identify multi-application congestion in the queue of grid-connection applications.

Disclosing available hosting capacity helps data center developers in evaluating potential sites for efficient planning and development. Available capacity information should be kept continuously up to date. DSO Entity and ENTSO-E formed the European Grid Action Plan in 2023 to improve visibility on available grid capacity, helping developers to plan renewable supply and electrified demand projects across the EU \cite{DSO_Entiry}. Some TSOs and DSOs have disclosed host capacity for demand connection \cite{Energinet_Map, TenneT_Map, Ireland_DSO_Map, NG_Map}.
As previously reported, there are concerns that data centers' grid connections will lead to higher electricity rates for residential customers. As one measure to address this, several TSOs are considering or implementing a cost-allocation system in which a portion of the grid reinforcement fees associated with grid connections is paid by data center operators. Starting in April 2023, the UK applied a Demand High-Cost cap under which, if the reinforcement costs incurred when connecting a customer to the distribution system exceed £1,720/kVA, the customer is required to bear the excess amount \cite{Ofgem_HCC}.

Fast-track under the Connect and Manage policy has been planned by PJM, and the concept was published in Jan. 2026. It offers the Critical Issue Fast Path, which will serve as an alternate path for any entity seeking to mitigate curtailment risk by bringing their own generation to the system, such as data centers that operate their own power plants. PJM also introduced the "Connect and Manage" program to large-scale demand projects that do not involve new power plants, allowing them to connect without waiting for grid upgrades \cite{PJM_BYONG}. EirGrid also began requiring data centers of 10~MVA or more to bring on-site or proximate generation since Dec 2025 \cite{EirGrid_LEU}.

Key technical requirements for data centers, including fault ride-through (FRT), power quality, and ramping limitations, should be considered grid codes. While ENTSO-E published recommendations of demand connection codes (DCC) requirements for large capacity loads \cite{ENTSO-E_Rec}, European TSOs have discussed and applied them into their grid codes \cite{Energinet_FRT, RTE_FRT, EirGrid_FRT, Tennet_DCC}. Each requirement varies depending on power system capacity and other conditions. For example, AESO strictly limits ramping to 10~MW/min, while Energinet allows up to 60~MW/min \cite{ESIG}.

Based on the grid codes and the extent to which relevant regulations have been implemented, all TSOs are either implementing or considering measures in response to the increase in requests for data center connections. As of the time of this paper’s submission, Canada, Denmark, and the Netherlands have paused the acceptance of new connection applications. In grids with a small original capacity, the tendency of data center connections to increase their share of demand also contributes to the suspension of new connections.

When comparing the status of discussions by region, similarities can be seen in the items being addressed or considered across regions such as North America and Europe. In North America, the North American Electric Reliability Corporation (NERC) is responsible for establishing uniform regulations and technical standards, while in Europe, this role is fulfilled by ENTSO-E. Because it is relatively easy to generalize and roll out technical standards developed and implemented in regions with early adoption, the items under revision or consideration in the grid codes of North America and Europe are becoming increasingly similar. In fact, ENTSO-E issued a recommendation to each TSO in 2025 regarding revisions to the Demand Code, which includes data centers. Although Japan has an organization called OCCTO responsible for formulating grid operation rules for Japan’s TSOs, discussions on revising the grid code are still in the early stages because Japan has integrated fewer data centers into their power systems than North America and Europe.

In all of the regions compared, mechanisms to encourage demand response (DR), either through mandatory curtailment for data centers with non-firm connections \cite{PJM_Curtail, ERCOT_Curtail, AESO_Curtail, NESO_Curtail, TenneT_nonFirm, monterde2025non} or through voluntary participation in DR programs \cite{DCFlex}, have been introduced or are under consideration. The following section explains how data centers can provide the flexibility needed to adjust to demand.

\begin{table*}[t]
    \centering
    \renewcommand{\arraystretch}{1.2}
    \caption{Grid Codes and Related Regulations for Large-Load Connection \cite{UK_GridCode, German_GridCode, ERCOT_GridCode, Japan_Penalty, Ofgem_HCC, Dutch_GridCode, PJM_BYONG, ENTSO-E_Rec, Energinet_Map, TenneT_Map, Ireland_DSO_Map, DSO_Entiry, NG_Map, Energinet_FRT, RTE_FRT, EirGrid_FRT, ESIG, Tennet_DCC, EirGrid_LEU, PJM_Curtail, ERCOT_Curtail, AESO_Curtail, NESO_Curtail, TenneT_nonFirm, monterde2025non, DCFlex}}
    \label{tab:gridcodes}
    \vspace{-6pt}
    \begin{adjustbox}{width=\textwidth,center}
    \begin{tabular}{l|c|c|c|c|c|c|c|c|c}
    \hline \hline
    \textbf{System Operator} & \textbf{PJM} & \textbf{ERCOT} & \textbf{AESO} & \textbf{Energinet} & \textbf{TenneT} & \textbf{EirGrid} & \textbf{RTE} & \textbf{NESO} & \textbf{TEPCO}
    \\  
    Country & US & US & CA & DK & NL & IE & FR & UK & JP \\ \hline
    Currently Pausing New Applications & $\times$ & $\times$ & $\circ$ & $\circ$ & $\circ$ & $\times$ (done) & $\times$ & $\times$ (done) & $\times$ \\ \hline
    First-ready, First-served & $\circ$ & $\circ$ & $\times$ & $\triangle$ & $\times$ (\textreferencemark 1) & $\times$ (\textreferencemark 2) & $\circ$ & $\circ$ &$\times$\\ \hline
    Penalty for Delayed Project & $\circ$ & $\circ$ & $\circ$ & - & - & - & - & $\circ$ & $\triangle$\\ \hline
    Screening Multi-application & $\triangle$ & $\circ$ & - & - & - & - & - & - & $\triangle$ \\ \hline
    Transparency on Available Demand Hosting Capacity & $\times$ & $\times$ & $\times$ & $\triangle$ (\textreferencemark 3) & $\circ$ (\textreferencemark 4) & $\circ$ (\textreferencemark 4) & $\triangle$ (\textreferencemark 4) & $\circ$ & $\circ$\\ \hline
    Cost Allocation of Grid Enhancement for Connection & $\triangle$ & $\circ$ & $\times$ & $\circ$ & $\triangle$ & $\times$ & $\circ$ & $\circ$ &$\times$\\ \hline
    Prioritizing Queue by Bringing Own On-site Power & $\triangle$ & $\circ$ & $\times$ & $\times$ & $\times$ & $\circ$ & $\times$ & $\times$ & $\times$\\ \hline
    Non-firm Connection or Curtailment Agreement & $\triangle$ (\textreferencemark 5) & $\circ$ (\textreferencemark 5) & $\circ$ & $\triangle$ & $\circ$ & $\circ$ & $\circ$ (\textreferencemark 5) & $\circ$ &$\triangle$ \\ \hline
    Fault Ride Through (FRT) Requirement & $\times$ & $\circ$ & $\circ$ & $\circ$ & $\circ$ & $\triangle$ & $\circ$& $\times$ & $\times$\\ \hline
    Power Quality Requirement & $\times$ & $\circ$ & $\circ$ &$\circ$ & $\circ$ & $\triangle$ & $\circ$ & $\times$  &$\times$\\ \hline
    Ramping Limitation (MW/min)& $\times$ & - & $\circ$ (10) & $\circ$ (60) & - & - & - & $\times$ &$\times$\\  \hline
    \hline
    \multicolumn{10}{l}{}\\ [-9pt]
    \multicolumn{10}{l}{Country Codes: US---United States, CA---Canada, DK---Denmark, NL---Netherlands, IE---Ireland, FR---France, UK---United Kingdom, JP---Japan} \\
    \multicolumn{10}{l}{Status: $\circ$: Applied; $\triangle$: Planning/Discussing; $\times$: Not Applied; -: Not Specified.} \\
    \multicolumn{10}{l}{(\textreferencemark 1): Prioritizing social importance of demand facility rather than readiness.} \\
    \multicolumn{10}{l}{(\textreferencemark 2): Restricting area and on-site generation instead of introducing first-ready, first-served policy.} \\
    \multicolumn{10}{l}{(\textreferencemark 3): Last updated in 2023.} \\
    \multicolumn{10}{l}{(\textreferencemark 4): Their DSOs provide available capacity heatmaps at the distribution level.} \\ 
    \multicolumn{10}{l}{(\textreferencemark 5): Data centers in those regions can join optional Demand Response (DR) programs \cite{DCFlex}.}
    \end{tabular}
    \end{adjustbox}
\end{table*}

\subsubsection{Unlocking Flexibility Capability of Data Centers by Components}
Various studies have been conducted on leveraging the flexibility of data center demand \cite{Adam_2014, Senga_2026, Takci_2025, ENTSOE2026DataCentres, Basmadjian_2017, Wang_2025, Colangelo_2026}. Because data centers have large loads, their power shaving provides flexibility in power and energy. Also, data center load can be shifted in time, providing temporal flexibility, if it is not on-demand task such as pre-training for AI models. It can even provide spatial flexibility by allocating its own tasks to geographically distant data centers \cite{Adam_2014}.
Flexibility from data centers can provide power, energy, temporal, and spatial flexibility by control their power demand \cite{Senga_2026}.
These flexibilities in data centers are provided by hardware components, such as on-site generation and cooling load, and IT load determined by operation algorithms \cite{ENTSOE2026DataCentres, Takci_2025}. 
IT load has constraints based on the type of computing task and service-level agreement (SLA), which guarantees performance and supports data centers' response times to their customers \cite{Basmadjian_2017}. Higher SLA limits the flexibility of data centers in ensuring their service levels \cite{Wang_2025, Colangelo_2026}.

\begin{figure}[t]
    \centering
    \includegraphics[width=1\linewidth]{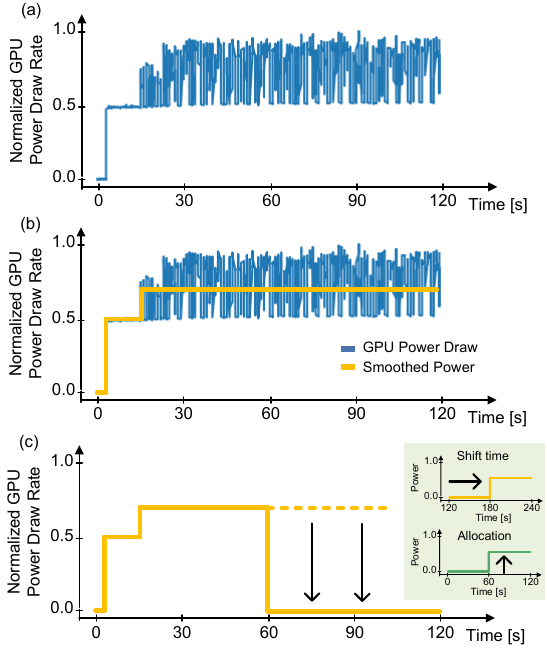}
    \caption{Needs of flexibility of data center. (a) Typical computing power flow of GPU \cite{MIT_DC}, (b) Needs of power smoothing by the data center components \cite{nvidia_smoothing}, (c) Needs of peak cut by shifting time, or allocating workload to another data center.}
    \label{fig:flex}
\end{figure}

Fig.~\ref{fig:flex} illustrates examples of the flexibility required of data centers. Fig.~\ref{fig:flex} (a) shows an example of a GPU power consumption curve taken from the MIT Supercloud Dataset \cite{MIT_DC}. The power output of this GPU begins to fluctuate around the 15-second mark. Similarly, according to NVIDIA, thousands of GPUs operate in lockstep and perform the same computation on different data during AI training, resulting in grid-level power fluctuations without power smoothing \cite{nvidia_smoothing}. As shown in Fig.~\ref{fig:flex} (b), data centers are required to smooth out this fluctuation and prevent adverse effects on the connected power grid \cite{Choukse2025PowerStabilization}. 
When grid congestion occurs, data centers must reduce their output. Possible methods for achieving this include "load shifting", which defers the use of the data center’s computing resources to a later time, as shown in Fig.~\ref{fig:flex} (c), or allocating computing resources to another data center. The former is an example of utilizing "Temporal Flexibility", while the latter is an example of utilizing "Spatial Flexibility". For example, the Google data center pilot project demonstrates demand-response capability by distributing IT load across different times and locations based on grid conditions \cite{Google_2023}.

According to \cite{ENTSOE2026DataCentres}, data centers comprise assets that provide flexibility, including DERs such as diesel, PV, and Wind generation; cooling load; IT load; and data center components such as Uninterruptible Power Supply (UPS) batteries. 
Data centers include on-site DERs for their primary power sources and for backup use. The most common form of on-site renewable energy generation for data centers is solar power, though there are also cases where other sources, such as micro wind turbines, fuel cells, biogas, and hydroelectric power, are used \cite{khosravi2024review}. Diesel generators are commonly used as backup power sources. Gas turbines and Fuel cells have been used as a cleaner alternative in some cases, and batteries have been considered as another backup source \cite{kambhampati2024moving}.
The cooling load refers to the electrical load consumed by the cooling system in a data center. In water-cooled systems, chilled water generated by chillers and cooling towers is distributed to each rack in the data center via distribution pumps to provide cooling \cite{dayarathna2015data}. Cooling systems integrated with thermal energy storage (TES) for data centers can reduce power and energy use and shift capacity over hours \cite{zhang2025unlocking}. As TES provides response times from minutes to hours, it is not suitable for frequency regulation services.
UPS systems provide critical backup power to operating data centers during grid failures. They provide instantaneous power on demand through their batteries and bridge the gap until other backup sources can provide it. A study proposes utilizing the energy flexibility of UPS batteries, as they have been used for less than 0.1\% of their operational lifespan \cite{Lunlong_2026}. 

Research in \cite{Gyang_2026} compared power stabilization methods for AI data center loads, using on-site assets including STATCOM, BESS under grid-forming control (GFM-BESS), and a combined approach using GFM-BESS and UPS.
STATCOM provides voltage stabilization, smooths load variations, and improves power quality at the Point of Common Coupling (PCC). GFM-BESS offers the additional advantage of providing greater energy flexibility. The study mentioned that operating a UPS to mitigate load fluctuations in AI workloads can achieve power smoothing cost-effectively. 

There have been cases where the IT load profiles and flexibility potential of AI data centers were studied in advance.
Traditional data centers process different types of tasks asynchronously and have a baseload consumption profile with minimal load fluctuations. AI data centers are expected to increasingly perform tasks such as model training, inference, and fine-tuning in a centralized manner within a single data center. Consequently, the distinct power consumption profiles of GPUs for each task are more likely to be reflected in the overall data center load profile \cite{nvidia_smoothing}. 
The load profile for AI training involves significant power consumption ramping. Furthermore, it has been reported that inference, such as that performed with large language models, exhibits high volatility in load profiles \cite{Li2024unseen}.
The study demonstrated a result of temporal flexibility utilization that applying workload orchestration to a production-scale GPU cluster of 130~kW enables the system to handle multiple DR events—including real-time grid dispatch, sustained power curtailment, carbon-aware computations, and geographically distributed load shifting—while maintaining service levels for AI priority training and inference jobs \cite{Williams2026power}. 

\begin{figure}[t]
    \centering
    \includegraphics[width=0.95\linewidth]{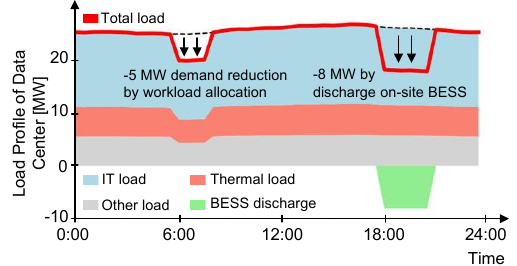}
    \caption{Example of flexibility utilization by data center; Load reduction by workload allocation between 6:00-8:00 and by discharging on-site BESS between 18:00-21:00.}
    \label{fig:available}
\end{figure}

Fig.~\ref{fig:available} shows an example of a data center flexibility utilization case. The data center load profile is based on publicly available 30-minute normalized data by UK Power Networks; specifically, the profile for a single data center connected to the extra-high-voltage grid on April 1, 2026 \cite{UKPN}, has been scaled up assuming a capacity of 60 MW. 
Data center demand includes IT load, thermal load, and other loads such as power supply losses. The share of each load (IT: 56\%, thermal: 22\%, and other loads, such as power supply losses: 22\%) is assumed based on the breakdown of energy consumption for colocation data centers reported in IEA \cite{IEA_AI_Demand}.
Fig.~\ref{fig:available} assumes that two DR events were triggered on the same day: the first DR, from 6:00 to 8:00, shows a case where the data center’s own demand was reduced through workload allocation, while the second DR, from 18:00 to 21:00, shows a case where the net load at the data center’s connection point was reduced by discharging the on-site BESS. The appropriate source of flexibility will vary depending on the time of day of the DR event, SLA constraints, the importance of the data center’s own resource tasks during the DR period, and the operational status of other data centers.

\subsubsection{Unlocking Spatial Flexibility Capability of Data Centers over Regions}
Because data centers are widely distributed across cities, countries, and continents, they can provide spatial flexibility. Spatial flexibility use cases can be mainly categorized into two types: local grid congestion mitigation and global synchronous grid balancing. In the Eastern Japan Synchronous System, which includes Tokyo, there were instances of supply constraints during peak periods, such as in June 2022 \cite{OCCTO_2022}. In such situations, data centers can allocate their load to another synchronous system even if there is no interconnection between the two synchronous grids. The concept of virtually interchanging power and energy between data centers, not through power lines but through fiber communication networks, has been called "virtual interconnector \cite{Kelly_2016}" or "virtual power lines \cite{weng2023distributed}." 
Attempts at resource allocation between data centers on different continents (synchronous areas), as described in \cite{Google_2023}, are expected to help reduce grid congestion and prevent blackouts. Research in \cite{Senga_2026} shows the potential to reduce data center operational costs by leveraging spatial flexibility across three interconnection areas in the US. Based on these concepts, this paper discusses the potential for further utilization of spatial flexibility, extending from synchronous power systems to intercontinental scales.

\begin{table*}[t]
    \centering
    \renewcommand{\arraystretch}{1.2}
    \caption{Global Synchronous Grid Area that Contains Major Data Center Corridors, Peak-Load Capacity, and DC Capacity \cite{IEA_AI, NERC_2025, NESO_2025, ENTSOE_PeakDemand, EMBER_China, OCCTO, Korea, India, AEMO_PeakDemand, EMI_2024, Eirgrid_2024, SONI_2025, DEA_PeakDemand}}
    \label{tab:spatial}
    \vspace{-6pt}
    \begin{adjustbox}{width=\textwidth,center}
    \begin{tabular}{c|c|c|c|c}
    \hline \hline
    \multirow{2}{*}{\textbf{Synchronous Area}}\textbf{} & \multirow{2}{*}{\textbf{Geographical Location}} & \textbf{Peak-Load Capacity} & \multirow{2}{*}{\textbf{Main Data Center Hub}} & \textbf{Operational Data Center} \\
     &  & \textbf{in 2024 [GW]} & & \textbf{Capacity in 2024 [GW] } \\ \hline
    Eastern Interconnection & Eastern area of USA and Canada & 621  (\textreferencemark 1) & Virginia, Atlanta & \multirow{3}{*}{53.7} \\ \cline{1-4}
    Western Interconnection & Western area of USA and Canada &  171  (\textreferencemark 1) & Phoenix, Alberta & \\ \cline{1-4}
    Texas Interconnection & Texas, USA & 85  & Dallas & \\ \hline
    RG Great Britain & UK Great Britain & 44.2 & London & 2.6 \\ \hline
    RG Ireland & Ireland + Northern Ireland & 7.1 (\textreferencemark 1,2) & Dublin & \multirow{3}{*}{11.9} \\ \cline{1-4}
    RG Continental Europe & Europe excluding Nordics & 387  (\textreferencemark 3)& Amsterdam, Frankfurt & \\ \cline{1-4}
    RG Nordics & Nordics & 52.7  (\textreferencemark 3) & Stockholm & \\ \hline
    State Grid Corporation of China & China mainland excluding south & \multirow{2}{*}{1465}& Shanghai, Beijing, Sichuan & \multirow{2}{*}{31.9} \\ \cline{1-2} \cline{4-4}
    China Southern Power Grid & Southern China mainland &  &  Pearl River Delta, Guiyang & \\ \hline
    Hokkaido synchronous area & Northern Japan & 4.4 & Sapporo & \multirow{4}{*}{6.6} \\ \cline{1-4}
    East Japan synchronous area & Eastern Japan & 70.6  (\textreferencemark 1) & Tokyo & \\ \cline{1-4}
    West Japan synchronous area & Western Japan & 90.8  (\textreferencemark 1) & Osaka & \\ \cline{1-4}
    South Korea synchronous area & South Korea & 97.1  & Seoul & \\ \hline
    India synchronous area & India & 250  & Mumbai & 3.6 \\ \hline
    National Electricity Market & Eastern Australia & 33.0 & Sydney & \multirow{4}{*}{1.6} \\ \cline{1-4}
    South West Interconnected System & Western Australia & 4.2  & Perth &  \\ \cline{1-4}
    North New Zealand & North New Zealand & \multirow{2}{*}{7.0} & Auckland &  \\ \cline{1-2} \cline{4-4}
    South New Zealand & South New Zealand &  & Christchurch &  \\ \hline
    \hline
    \multicolumn{5}{l}{}\\ [-9pt]
    \multicolumn{5}{l}{(\textreferencemark 1): Peak load is calculated by summing up peak load in each subregion, which may contain different time.} \\
    \multicolumn{5}{l}{(\textreferencemark 2) Sum of peak demand by Ireland of 5.6~GW \cite{Eirgrid_2024} and Northern Ireland of 1.5~GW \cite{SONI_2025}.} \\
    \multicolumn{5}{l}{(\textreferencemark 3): RG Nordics contains Norway, Sweden, Finland, and the DK2 region of Denmark, assuming 40\% of Danish peak demand is accounted for based on \cite{DEA_PeakDemand}.} \\
    \multicolumn{5}{l}{RG Continental Europe contains the rest of the areas excluding Great Britain, Ireland, and Georgia, and including DK1, as 60\% of Danish peak demand \cite{DEA_PeakDemand}.}
    \end{tabular} 
    \end{adjustbox}
\end{table*}

Table~\ref{tab:spatial} shows the synchronous systems to which major data center deployment regions belong, along with the peak load capacity in 2024 \cite{NERC_2025, NESO_2025, ENTSOE_PeakDemand, EMBER_China, OCCTO, Korea, India, AEMO_PeakDemand, EMI_2024, Eirgrid_2024, SONI_2025, DEA_PeakDemand} and data center capacity deployed in each region as of 2024 \cite{IEA_AI}. The United States has the largest data center capacity, at 53.7~GW, across the Eastern Interconnection, Texas Interconnection, and Western Interconnection. In Europe, the United Kingdom alone has 2.6~GW of data centers. The UK is an independent synchronous system connected via HVDC to the Irish, Continental, and Nordic synchronous systems. Europe has 11.9~GW of operational data centers, most of which are located within the Continental European grid. While the Irish grid has a capacity of approximately 8~GW and peak demand reached 7.1~GW in 2024, it hosts 1.26~GW of operational data centers, meaning data center demand accounts for a large share of its total capacity. In Asia, 31.9~GW of data center capacity has been deployed in China, with the majority located within the synchronous grid managed by the State Grid Corporation of China. Southern China features a separate synchronous grid managed by China Southern Power Grid, which includes data center hubs such as the Pearl River Delta and Guiyang. South Korea and Japan together host 6.6~GW of data centers. South Korea operates as a single, independent synchronous grid. Japan primarily consists of three synchronous systems, Hokkaido, Eastern Japan, and Western Japan, which are connected via HVDC. India has a single, independent synchronous system that includes 3.6~GW of data centers. Australia and New Zealand have 1.6~GW of data centers in operation, the majority of which belong to the National Electricity Market (NEM) synchronous system. By allocating resources across these synchronous systems, it is possible to alleviate supply-demand imbalances that are difficult to address within a single synchronous system alone, thereby overcoming the constraints of HVDC interconnection. Since some tasks cannot handle data from other regions due to data latency and privacy constraints, it is also necessary to quantify which tasks can be flexibly allocated across regions.

\subsubsection{Incentivization of Flexibility Utilization from Data Center}
From the data center operator's perspective, the value of flexibility must exceed the value of Compute Load [\$/MWh] (VoCL), which translates cloud market economics into an implied willingness to pay for electricity. It has been reported that VoCL varies with the type of computing task and the class of AI \cite{Billimoria_2026}. On-demand tasks running within the scope of an SLA have the highest VoCL, making them difficult to utilize for flexibility. Interruptible tasks, such as model pre-training and computations involving checkpoints, have a VoCL less than one-tenth that of on-demand tasks subject to SLAs, indicating they are easier to use for flexibility. PJM and European TSOs are considering a transition to connect \& manage for large-scale demands such as data centers \cite{PJM_BYONG}. Since the value of available flexibility varies by AI class and task, it is important to identify the flexibility that data centers can naturally contribute and to establish a clear division of roles, such as requiring data centers to provide electricity price-hedging functions based on VoCL price tiers or offering priority service contracts.

\subsection{Observations and Recommendations}

Through the widespread occurrence of extended lead times in response to the growing need for integration into today’s data center power systems, we identified the bottlenecks in the supply chain, integration processes, and power system operations that contribute to these longer lead times. It was observed that the processes on the power system operations side tend to span a longer period than data center construction.
The analysis also covered the implementation status and plans for grid codes and related regulatory reforms being carried out by TSOs in major data center hub regions to resolve these operational bottlenecks. While efforts to establish unified standards are underway in both North America and Europe, there are certain differences in the areas being revised: in Europe, efforts to increase transparency regarding available hosting capacity are progressing, whereas in North America, discussions are advancing on a policy allowing data centers to bring their own on-site power. Mechanisms are being prepared or considered to require data centers to adjust their demand through various means, such as requesting curtailment via non-firm connections or offering voluntary DR programs.

Data centers must smooth instantaneous load fluctuations, provide temporal flexibility to shift load over time, and offer spatial flexibility to reduce demand by allocating workloads to other data centers. The trend in providing flexibility has been introduced using DERs on-site, thermal energy systems for cooling, and IT workloads. Because the allocation capability of IT workloads varies with task priority and SLA constraints, the resources available for flexibility vary by time of day.

Additionally, it was discussed how the spatial flexibility of data centers can mitigate not only local grid congestion but also supply-demand imbalances in global synchronous systems. Peak demand levels in synchronous systems and data center capacity in areas with major data center hubs were presented to identify potential sites for virtual interconnectors (virtual power lines) across regions, beyond power transmission via HVDC or international interconnectors. VoCL was introduced as a potential metric to motivate data center DR. Further refining this concept will clarify the flexibility available for virtual interconnectors across regions.

\section{Powering the Grid-to-Chip Voltage Cascades: Power Electronics Enablers}

\subsection{From Grid Boundary to Chips: A Cross-Scale Hierarchy}

\begin{figure}[t]
    \centering
    \includegraphics[width=\linewidth]{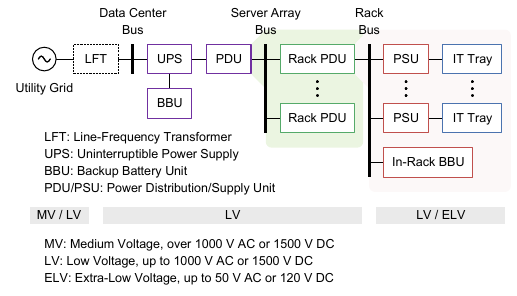}
    \caption{General grid-to-chip hierarchy of data center power supply supported by power electronics interfaces \cite{ornl2026, ocp2026lvdc}.}
    \label{fig_dc_hierarchy}
\end{figure}
\begin{figure}[t]
    \centering
    \includegraphics[width=\linewidth]{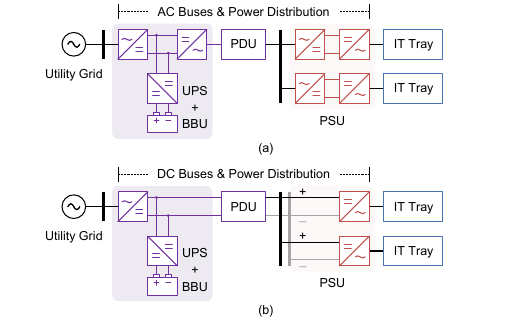}
    \caption{Transitioning from (a) AC to (b) DC facility buses, where fewer conversion stages bring about higher energy efficiency and reduced installation and maintenance costs.}
    \label{fig_dc_hierarchy_2}
\end{figure}

The power path in an AI data center is typically a continuously voltage-decreasing, current-increasing cascading hierarchy, as illustrated by the grid-to-chip hierarchy in Fig.~\ref{fig_dc_hierarchy} \cite{ornl2026, ocp2026lvdc}: voltage is stepped down from medium/low-voltage (MV/LV) AC grids towards on-tray extra-low voltage (ELV) level, where online uninterruptible power supplies (UPS) provide backup support for rack-level buses (conventionally 48~V DC for information-technology (IT) trays) \cite{krein2017}. The voltage-class labels in Fig.~\ref{fig_dc_hierarchy} are worth foregrounding given that they mark where this hierarchical shift crosses a regulatory boundary more than merely physical one.

Power electronics underpins all stages of this hierarchy, performing both AC-DC conversion, voltage stepping and desired power distribution with power semiconductors at high controllability and power efficiency. Technically, each tier in Fig.~\ref{fig_dc_hierarchy} does not share identical design objectives despite belonging to the same power delivery chain. The baseline was designed for a load profile that has however changed significantly throughout the decade, while as individual rack power already scales towards megawatt level or even higher \cite{pcim2025path}, the acuteness of high-power and high-volatility AI data center loads is no longer negligible \cite{Choukse2025PowerStabilization}: a 1~MW rack corresponds to approximately 20~kA of current on a conventional 48~V ELV bus, that yields in-rack bus design extremely challenging. This fundamental constraint in current escalation---rather than efficiency alone---is driving the ongoing paradigm shift in data center power electronics. Beyond backup duty, they are also increasingly called upon to participate in grid-interfacing power regulation or management. Those transitions are inspiring new power electronics designs at the rack and tray level, yet dedicated standardization for LV IT trays remains an open problem \cite{shaukat2026pcimpanel}.

The major design priorities can be summarized as Table~\ref{tab_pe_design}, while they are supposed to serve for the following aspects targeting distinct physical evolutions \cite{ornl2026, kolar2026sst, nvidia800vdc1}:
\begin{enumerate}
    \item \textit{From the 48~V in-rack ELV DC bus to 800~V LV DC bus}. As current scales inversely with voltage at fixed power, ELV naturally pushes the conduction loss as well as conductor cross-section size to their engineering limits \cite{kolar2026sst}. An example is reported by NVIDIA where adopting 800~V DC in place of 415~V AC reduces the copper requirements by 45\% \cite{scalia2026bodo, nvidia800vdc2}, directly relieving the 20~kA bottleneck as identified, whereas future possibility is also seen towards 1500~V DC ($\pm$750~V) to fully utilize the standardized LV DC framework \cite{ocp2026lvdc}.

    This redefines where the DC-DC conversion primarily occurs, formulates 800~V in-rack DC distribution, and shifts the voltage step-down towards the IT trays \cite{nvidia800vdc1, nvidia800vdc2, infineon800v}. However, it also brings along a 16.7\texttimes\ jump of the conversion ratio compared with conventional 48~V cases that challenges conventional small-scale DC-DC conversion design. To this end, new topology families while preserving power density and efficiency are therefore motivated as an important technical focus.

    \item \textit{From LV AC to LV DC facility buses}. Since IT loads are inherently DC at the point of consumption, conversion into intermediate AC buses shows limited advantages other than conforming to the legacy AC infrastructure, as illustrated in Fig.~\ref{fig_dc_hierarchy_2}. DC power distribution improves the power efficiency in distribution bus bars and reduces the conductor volume \cite{kolar2026sst}, meanwhile eliminating the redundant stages in the power chain improves operational reliability and further enables energy storage to directly interact with server rack behaviors and perform faster response to server-rack load transients. Nevertheless, with less fault-interruption ability, DC distribution grounding and protection must in turn be designed more rigorously than in conventional AC infrastructures.

    \item \textit{From line-frequency transformer (LFT) to solid-state transformers (SSTs)}. This shift is grounded in the transformer scaling physics, where, by replacing the bulky 50/60~Hz LFT cores with power electronics converter cells coupled through intermediate medium- to high-frequency transformers (MFTs/HFTs), the volume is significantly decreased and higher power density can be achieved \cite{kolar2026sst}. SSTs impart modularity and scalability that allow the power interfaces to be more readily reconfigured and keep pace with the requirements from rapidly evolving server racks, of which the standardization is also being defined that aims to highlight their responsibilities on both grid and data center sides \cite{ocp2026sst}.
\end{enumerate}

\begin{table}[t]
    \centering
    \renewcommand{\arraystretch}{1.2}
    \caption{Design Priorities for Power Electronics in Data Centers:\\Key Features and Justifications}
    \vspace{-6pt}
    \label{tab_pe_design}
    \begin{tabularx}{\linewidth}{>{\raggedright\arraybackslash}p{0.25\linewidth}X}
    \hline\hline
    \textbf{Priority} & \textbf{Technical Justification} \\ \hline
    
    Voltage/current staging & Cascaded power conversion from MV/kV-level utility feeds to sub-1~V chip rails, where step-down ratio grows sharply shaping the choice of converter topologies. \\ \hline
    
    Modularity & Scalable modification of rails, racks or converter cells with least interruption on the power delivery of IT trays; however benefits are bounded by the modularization penalty in terms of per-cell isolation requirements. \\ \hline
    
    Capability of power fluctuation, harmonics and power quality & Large and synchronized AI server power swings inject harmonic content that risks exciting resonant modes in upstream grids, also motivating active power-factor correction (PFCs) or filtering at the grid interface. \\ \hline
    
    Fast dynamic response & Sub-millisecond compute or communication phase transitions in AI computing tasks demand high converter control bandwidth and low output impedance to preserve voltage tracking at the point of load. \\ \hline
    
    Power efficiency and density & Minimizing power loss and volume simultaneously at the Gigawatt-scale facility power benefits cost optimization as well as offsetting the overhead of cooling design. \\ \hline
    
    Reliability or fault-tolerant architecture & Continuous operation under mission-critical services underlines fault avoidance, robustness against unexpected faults, or the ability to limit the impacts of any single failure on coupled server units. \\ \hline
    
    Isolation and protection & Galvanic isolation shifts from optional (ELV) to mandatory (LV/MV); DC distribution inside data centers further raises concerns about protection designs. \\ \hline
    
    EMI/EMC performance & High $\mathrm{di/dt}$ or $\mathrm{dv/dt}$ from fast high-frequency switching must avoid coupling into co-located and EMI-sensitive xPU and memories. \\ \hline
    
    Thermal/cooling design & Loss not delivered as useful IT load power is dissipated as heat that affects both efficiency and reliability. Higher power density of power electronics and compact converter packaging require synthesized thermal validation and optimization. \\ \hline
    \hline
    \end{tabularx}
\end{table}

\subsection{Architectures of Power Electronics Converters}

\begin{figure}[t]
    \centering
    \includegraphics[width=\linewidth]{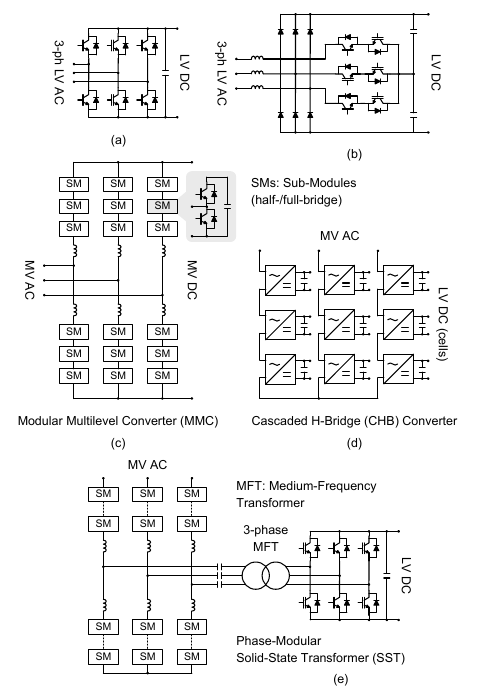}
    \caption{Representative AC-DC power electronics converter topologies for data center applications \cite{kolar2026sst, marco2016sst}, including: (a) two-level converter; (b) Vienna rectifier; (c) modular multilevel converter (MMC); (d) cascaded H-bridge (CHB) converter; and (e) phase-modular solid-state transformer (SST) architecture.}
    \label{fig_topology_dcac}
\end{figure}

\begin{figure}[t]
    \centering
    \includegraphics[width=\linewidth]{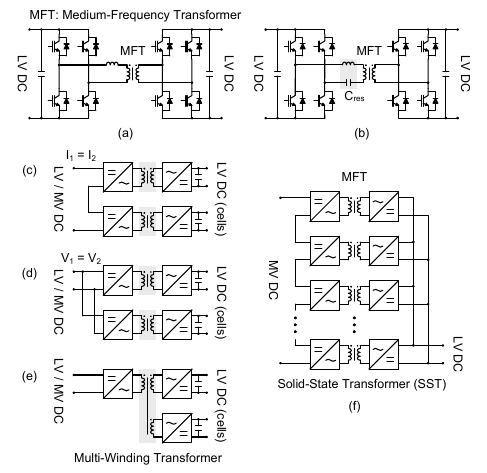}
    \caption{Representative DC-DC power electronics converter topologies for data center applications \cite{ornl2026, marco2016sst}, including: (a) dual active bridge (DAB) converter; (b) LLC resonant converter; (c)-(e) voltage stepping configurations based on (c) primary-side voltage sharing, (d) primary-side current sharing, and (e) power distribution through a multi-winding transformer; and (f) solid-state transformer (SST).}
    \label{fig_topology_dcdc}
\end{figure}

Various architectures of power electronics converters enable the aforementioned power conversion, voltage stepping and power distribution at each bus hierarchy, with representative AC-DC and DC-DC topologies summarized in Fig.~\ref{fig_topology_dcac} and Fig.~\ref{fig_topology_dcdc}, respectively \cite{ornl2026, kolar2026sst, marco2016sst}. The primary AC-DC building blocks are typically the two-level bridge shown in Fig.~\ref{fig_topology_dcac}(a) or the Vienna rectifier structure in Fig.~\ref{fig_topology_dcac}(b), with the possibility of extension in voltage levels and plus the initiative of using MFTs/HFTs for both galvanic isolation and higher power density. At the grid-interfacing data center bus, this shares common operational philosophy with the general SSTs for energy routing \cite{marco2016sst, pes2025sst, sharida2025sstmag} but often without the stage of converting LV DC back to AC. DC-DC converters used for PDUs or UPS/backup-battery units (BBUs) are most commonly configured based on the dual-active bridge (DAB) or resonant converters (Fig.~\ref{fig_topology_dcdc}(a)(b)) where bi-directional power flow is especially preferred, while uni-directional topologies like forward, flyback and push-pull configurations are also reported in \cite{khan2024dcdc} that are more suitable for on-board power conversion within IT trays.

A common design principle for both families is the distribution of voltage and current stress across converter cells: connection in series for voltage division at MV buses and in parallel for current sharing at LV buses---typically forming the input-series-output-parallel (ISOP) configuration described by \cite{ornl2026}. The MV boundary is thereby reached by cascading lower-voltage sub-modules, as in modular multilevel converters (MMCs) (Fig.~\ref{fig_topology_dcac}(c)) or cascaded H-bridge (CHB) converters (Fig.~\ref{fig_topology_dcac}(d)), while load-side power distribution can be facilitated through common DC bus or multi-winding transformers. An example of the 800~V-to-1~V conversion is reported by \cite{palani2026dc800to1}, and \cite{kolar2026sst} underlines \textit{hybridization} (combination of LFTs and/or power electronics converters), \textit{modularization} (modularization of SST topologies), and \textit{decentralization} (power distribution at different hierarchies) for future data center power distribution architectures. Among them, converter modularity benefits the manufacturing and improves field-scalability and replaceability, but it comes alongside the fundamental trade-off concerning voltage/current capability, insulation requirements, maintenance costs and other embodied costs such as space occupation. A key point lies in the "modularization penalty" \cite{kolar2026sst}: each isolation barrier may still be subject to the full operating voltage rating, which bounds the converter cell counts, per-cell power rating and the reasonable power rating of SSTs.

\begin{figure}[t]
    \centering
    \includegraphics[width=\linewidth]{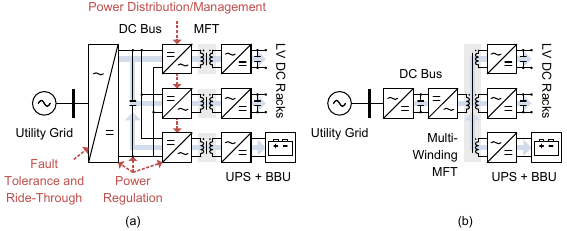}
    \caption{Technical perspectives on SSTs for data centers anticipating opportunities in (a) SSTs enabling power regulation, management and system resilience, and (b) multi-winding MFTs facilitating shorter power paths, higher power density and distinct power controllability.}
    \label{fig_sst_frontier}
\end{figure}

To be specific, several research frontiers and challenges have been particularly identified for SSTs as the leading interfacing technology for data centers incorporating the aforementioned requirements and possibilities (see Fig.~\ref{fig_sst_frontier}):

\begin{enumerate}
    \item \textit{Power distribution and management}. Unlike conventional LFTs, power electronics converters in SSTs offer considerable degrees of freedom in control. In SSTs with an ISOP configuration supplying server racks, this enables active power distribution and management, which allows power and energy flexibility of the system to be exploited for AI load scheduling and power routing. Leveraging this capability is a key to ensuring compliant grid integration and service quality under evolving AI workloads.

    \item \textit{Power regulation and system resilience}. On the load side, SSTs are primarily responsible for ensuring system resilience, where robustness, like capability of grid-side fault ride-through, should be guaranteed. On the grid side, as power electronics turn data centers into grid-interactive systems, energy storage in BBUs can also support power factor correction (PFC) or power regulation, with promising directions including grid-supportive energy routing and power smoothing to mitigate the oscillations from AI workloads \cite{Ko2026WideAreaOscillation}. This capability can further be promoted by MMC-based front ends \cite{marco2022mmcsst}.

    \item \textit{Exploitation of the MFTs/HFTs}. The transformers in SSTs, i.e., MFTs/HFTs, are key to reducing the PDU/PSU volumes and enhancing system power density. While power electronics converters dominate active power distribution, the magnetic coupling through the cores also supports power flow besides galvanic isolation. Multi-winding transformers (Fig.~\ref{fig_sst_frontier}(b)) shorten the power paths and enhance efficiency, and the inherent physical coupling and constraints also expedite partial power processing \cite{marco2021multiwinding}. Planar transformers are meanwhile emerging as an appealing approach for integrating the PDUs/PSUs into IT trays, promising benefits for both power density and integrated cooling design \cite{marco2026planar}, yet their applications remain to be matured to overcome the more pronounced impacts of on-board parasitics.
    
    \item \textit{Hardware optimization and reliability}. Hardware optimization of SSTs must account for reliability and electromagnetic compatibility (EMC) in addition to efficiency and power density. This concerns not only power semiconductors and capacitors, but equally the design of MFTs/HFTs. Besides, control could also contribute as complementary approaches, like active thermal control \cite{marco2016thermal}, where the controllability of modularized cells are harnessed to assist with maximizing the usage of hardware and achievable performances.
    
    \item \textit{Smart operation}. In view of these aspects, SSTs are therefore supposed to be smart \cite{marco2026pels}: Rather than exclusively pursuing efficiency or power quality, SSTs should comprehensively address a sufficient set of functionalities that goes beyond conventional fundamental design, including power controllability, stability, reliability or modularity, approaching the Pareto fronts among them. Also, the performance baselines should not be constant but should evolve alongside the future development of AI workloads and data centers.
\end{enumerate}
Other than the IEEE PES Task Force (TF) Paper \cite{pes2025sst}, two upcoming frontline benchmarks for SST design and integration are IEEE Standard P3105 \cite{ieeeP3105} and CIGRE WG B4.91 \cite{cigreB491}. Their expected scopes are also outlined and compared in \cite{marco2026pels}: Besides the common ground, IEEE PES TF additionally highlights the modeling through architectural and operational concepts, while CIGRE WG B4.91 and IEEE Std. P3105 are projected to formalize system-level assessment and hardware implementation, respectively.

\begin{figure}[t]
    \centering
    \includegraphics[width=\linewidth]{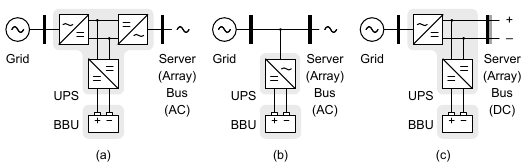}
    \caption{Interfacing of server-level backup battery units (BBUs) through power electronics uninterruptible power supply (UPS) systems, with (a) and (b) for AC server buses and (c) for DC ones.}
    \label{fig_dc_storage}
\end{figure}

When integrating the BBUs, the shift towards 800~V LV DC bus moves the chip-level loads progressively closer to the interconnection point, as exemplified in Fig.~\ref{fig_dc_storage}. The design is basically subject to power efficiency and operational durability, both of which generally benefit from minimizing the number of conversion stages between grid and load \cite{chen2023dcpower}---which becomes particularly outstanding given the tremendous power consumption of modern server loads \cite{Choukse2025PowerStabilization}. For modular topologies, the two considerations are further addressed by mitigating the unbalance across sub-modules and circulating currents between the phase legs.

UPS converters have also been developed from merely maintaining the output voltage towards grid-interactive assets. Commercially, bi-directional UPS systems for data centers are deployed to buffer the mutual impacts between the utility grid and AI servers, like grid-supporting \cite{ocp2026sst} and guaranteeing the power quality against AI workload fluctuations \cite{sina2026gridint}. Nevertheless, this should not be pursued simply by oversizing the storage capacity, but alternatively achieved by exploiting the flexibility from xPU loads---the alternation between near-TDP (thermal design power) and near-idle phases---to assist the power distribution.

Building on these observations, we further map the data center power supply architectures in \cite{ornl2026} throughout the hierarchies and summarize the observations into Table~\ref{tab_pwrarchmap}. Rather than mapping a single architecture, we aim to extend the same observations also to the emerging 800~V-DC-class solutions like \cite{nvidia800vdc2}, as a comprehensive reference whenever the relevant hierarchy levels or converter stages are involved.

\begin{table*}[t]
    \renewcommand{\arraystretch}{1.35}
    \renewcommand{\tabularxcolumn}[1]{m{#1}}
    \centering
    \caption{AI Data Center Power Supply Architecture---A Grid-to-Chip Hardware Map from \cite{ornl2026}}
    \label{tab_pwrarchmap}
    \vspace{-6 pt}

    \setlength{\tabcolsep}{3pt}

    \begin{tabularx}{\linewidth}{>{\raggedright\arraybackslash}m{0.11\linewidth}%
    >{\raggedright\arraybackslash}m{0.17\linewidth}%
    >{\raggedright\arraybackslash}m{0.42\linewidth}%
    >{\raggedright\arraybackslash}X}
        \hline\hline
        \textbf{Hierarchy} & \textbf{Hardware Proposition} & \textbf{Technical Profiles} & \textbf{Key Observations}\\ \hline
        
        Utility Level
        & LFT Grid Interface &%
        \textbf{Topology}: Passive transformer \linebreak
        \textbf{Objective}: Grid interfacing, galvanic isolation \linebreak
        \textbf{Voltage level}: MV AC to LV AC (480~V level, 3-phase)
        &%
        Substation-level insulation and grounding
        \\ \cline{2-4}
        & MV-SST Grid Interface &%
        \textbf{Topology}: AC-DC bi-directional, modularization for ISOP \linebreak
        \textbf{Objective}: Grid interfacing, galvanic isolation and grid-level interactivity \linebreak
        \textbf{Voltage level}: MV AC to LV DC (800~V level)
        &%
        Grid synchronization + PFC \linebreak
        Fast data-center-level protection and fault isolation \linebreak
        MV-rated device and insulation design
        \\ \cline{2-4}
        & Facility Front End \linebreak
        (Facility PDU) &%
        \textbf{Topology}: AC-DC-AC or AC-DC bi-directional \linebreak
        \textbf{Objective}: AC-synchronization + rack-level power distribution \linebreak
        \textbf{Voltage level}: LV AC to LV DC (800~V level)
        &%
        AC bus synchronization + PFC \linebreak
        Data-center-level protection
        \\ \cline{2-4}
        & Facility UPS Interface &%
        \textbf{Topology}: AC-DC or DC-DC bi-directional \linebreak
        \textbf{Objective}: Data-center-level energy storage + power stabilization, grid interactivity (power conditioning) \linebreak
        \textbf{Voltage level}: LV AC or DC to LV/ELV DC batteries
        &%
        Battery management \linebreak
        Hot-swap capability \linebreak
        (For DC buses) DC fault detection capability
        \\ \hline

        Rack Level
        & Rack-level "sidecar" PDU &%
        \textbf{Topology}: AC-DC, modularization for parallel racks/PSUs \linebreak
        \textbf{Objective}: AC-synchronization + rack-level power distribution \linebreak
        \textbf{Voltage level}: LV AC to LV/ELV DC (800~V/48~V level)
        &%
        AC bus synchronization + PFC \linebreak
        Data-center-level protection
        \\ \cline{2-4}
        & Rack PDU and PSU &%
        \textbf{Topology}: DC-DC, modularization for multiple IT trays \linebreak
        \textbf{Objective}: Voltage step-down for legacy IT-tray compatibility \linebreak
        \textbf{Voltage level}: LV DC to ELV DC (48~V level)
        &%
        High-current capability (kA level) \linebreak
        Rack-level interactivity \linebreak
        (For 800~V DC) High step-down ratio
        \\ \cline{2-4}
        & In-Rack BBU Interface &%
        \textbf{Topology}: DC-DC bi-directional, modularization for multiple BBU cells \linebreak
        \textbf{Objective}: Rack-level energy storage + power stabilization \linebreak
        \textbf{Voltage level}: LV/ELV DC to LV/ELV DC batteries
        &%
        Sub-ms transient response \linebreak
        Battery management \linebreak
        Hot-swap capability
        \\ \hline

        Chip Level
        & On-tray PSU and IBC &%
        \textbf{Topology}: DC-DC, typically uni-directional \linebreak
        \textbf{Objective}: Intermediate bus conversion \linebreak
        \textbf{Voltage level}: LV/ELV DC to ELV $\mathrm{V}_\mathrm{IB}$ (12~V level, 1.8-24~V)
        &%
        High current density \linebreak
        High-$\mathrm{dv/dt}$, $\mathrm{di/dt}$ capability and EMI performance \linebreak
        (For 800~V DC) High step-down ratio
        \\ \cline{2-4}
        & On-chip VR &%
        \textbf{Topology}: DC-DC, typically uni-directional \linebreak
        \textbf{Objective}: PoL voltage regulation \linebreak
        \textbf{Voltage level}: ELV $\mathrm{V}_\mathrm{IB}$ to chip level (1~V level, $<$5~V)
        &%
        Sub-µs transient response \linebreak
        Very high current density \linebreak
        High-$\mathrm{di/dt}$ capability and EMI performance
        \\
        
        \hline\hline
        \multicolumn{4}{l}{}\\ [-9pt]
        \multicolumn{4}{>{\raggedright\arraybackslash}p{0.99\linewidth}}{\textbf{Note}: PFC---Power-Factor Correction; VR---Voltage Regulator; IBC---(on-tray) Intermediate Bus Converter; $\mathrm{V}_\mathrm{IB}$---(on-tray) Intermediate-Bus (IB) voltage, typically 12~V level. Other acronyms are defined as per Fig.~\ref{fig_dc_hierarchy}.}
    \end{tabularx}
\end{table*}

\subsection{Evolution of Power Electronics Devices}

The capability of the converter topologies are also determined by the power semiconductor devices. Forefront advancements mainly come down to three interlocking aspects: device type, emerging materials and device packaging.

\begin{enumerate}
    \item \textit{Device type}. In general, insulated gate bipolar transistors (IGBTs), metal-oxide-semiconductor field-effect transistors (MOSFETs) and the newer high-electron-mobility transistor (HEMTs) are the three families that are currently best suited to data center applications. IGBTs typically serve for higher-voltage and moderate-frequency stages (at the legacy AC-DC front ends or UPS rectifiers), with around tens of kHz switching frequency. MOSFETs or HEMTs, on the other hand, dominate where faster switching and lower conduction loss matter more, like power-factor correction (PFC) and point-of-load voltage regulation modules (PoL VRMs) \cite{intal2026devicedc}, which is increasingly essential especially given the scale and rapid fluctuation of AI workloads. Controllable thyristors and other multi-kV devices, however, are largely absent from the rack itself, since modularized topologies already cover the voltage level needed.

    \item \textit{Emerging materials}. Silicon (Si) devices are an economical default when voltage and frequency requirements are modest, while wide-bandgap (WBG) materials like SiC and GaN are proved to be better high-voltage high-$\mathrm{dv/dt}$ solutions \cite{millan2014survey, zhang2025wbg, ravindran2025wbg}. Reported data-center benchmarks illustrate the scale of the gain from Si to WBG devices in both efficiency and power density \cite{zhang2025wbg}, and \cite{ravindran2025wbg} compares several WBG-based application examples. A further tier, ultrawide-bandgap (UWBG) materials like diamond, Ga\textsubscript{2}O\textsubscript{3} or AlN \cite{zhang2025wbg, ravindran2025wbg, qin2023}, promises still higher device performances but remain to be fully demonstrated in practice.

    \item \textit{Device packaging}. Packaging of power electronics devices sets the parasitic parameters (especially inductance) that limits the device switching speed and thus power efficiency. This is considerably more significant for WBG and UWBG devices with high $\mathrm{di/dt}$ and $\mathrm{dv/dt}$, and the resulting design trend is to preserve the fast-switching performance through low-inductance interconnects \cite{intal2026devicedc}, including co-packaging of gate drivers or integrated cooling. Infineon's .XT diffusion-soldering interconnect has been reported \cite{infineon2026xt}, which reduces the junction-to-case thermal resistance $\mathrm{R}_\mathrm{th,\:JC}$ by up to 25\% compared with the prior generation, indicating an improvement in the thermal performance and thus allowing for higher power density as needed for data centers.
\end{enumerate}

\subsection{Chip-Level Power Supply Innovations and Trends}

\begin{figure}[t]
    \centering
    \includegraphics[width=\linewidth]{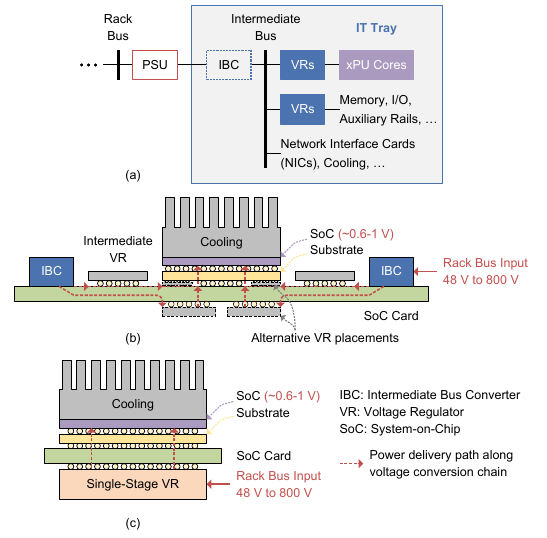}
    \vspace{-18pt}
    \caption{Representative chip-level power delivery architectures \cite{prakash2025vpd, zhu2026vpd, li2025vpd, ye2026chipsys}, (a) summary of the on-tray power delivery hierarchy, (b) lateral power delivery (LPD) where converters are placed beside the die, and (c) vertical power delivery (VPD) where converters are placed vertically underneath the die. Backside power delivery (BPD), in contrast, focuses on the semiconductor wafer \textit{within} the system-on-chip (SoC) and thus does not overlap with LPD or VPD.}
    \label{fig_chip_level}
\end{figure}

Besides the shift towards 800~V LV DC bus, the chip-level power delivery network (PDN) becomes a distinct focus to further enhance the power efficiency and density as AI workloads scale up. We summarize the on-tray power delivery hierarchy in Fig.~\ref{fig_chip_level}(a): With intermediate bus converters (IBCs) stepping the IT tray voltage input to intermediate buses, distributed voltage regulators (VRs) supply the xPU cores, memory, I/O, and auxiliary rails, meanwhile the on-tray ventilators for cooling and network interface cards (NICs) \cite{sohan2010nic} also consume a non-negligible share of power outside the VR branches.

The on-tray converter (both IBCs and VRs) substrates are integrated closer the xPU motherboard, aiming to address the "last inch" of power conversion from the converter (voltage regulator) input to the die and reduce the losses along the track paths (the IR drop). Chip manufacturers have been developing and commercializing the backside power delivery (BPD) at die level, where certain power-delivering interconnects are moved to the backside of the wafer beneath the transistors---as reported by Intel (PowerVia \cite{intel2023powervia1, intel2023powervia2}), TSMC (A16 \cite{tsmc2026a16}) and Samsung (SF2Z, \cite{samsung2026sf2z}). Though mature wafer fabrication process is demanded, it improves the logic density by freeing the front-side high-speed I/O routing next to the xPUs, and reduces the IR drop through widened power wiring.

Nevertheless, BPD only reaches the die's own internal metal stack, but external converters are still needed to feed the die, especially as the upstream bus voltage rises to 800~V where the insulation clearance required by the external converter becomes non-negligible. On the other hand, fully on-chip power supplies do not straightforwardly outperform that of off-chip ones at the same hierarchical scale \cite{wang2015onoffchip}, where the efficiency gained from proximity is offset by the on-die embedded passive components. The converters are conventionally placed beside the die, known as lateral power delivery (LPD), from which the vertical power delivery (VPD) architectures are investigated \cite{scalia2026bodo, prakash2025vpd, zhu2026vpd, li2025vpd, ye2026chipsys, infineon2025powering}, that relocate the converters vertically underneath the die and shorten the power delivery track paths---directly in contrast to the off-chip LPD solutions as illustrated in Fig.~\ref{fig_chip_level}(b) and (c). Relevant solutions have been reported by Google (patent, \cite{google2024}) and Infineon (OptiMOS power modules \cite{infineon2025optimos}). In essence, VPD goes beyond the electrical-dominated design and optimizes the board area around the processors by pushing the design also towards the mechanical boundaries---sparing space also for high-speed I/O around the xPUs, significantly benefiting the design for next-generation high-performance AI processors.

The said challenge in embedded passive components still exist for VPD, thus motivating the dedicated studies on planar transformers \cite{wen2026vpdtrafo} as well as cooling design \cite{choi2025vpdcooling} for VPD. Also, since BPD and VPD address different and non-overlapping segments of the chip-level power delivery paths, they are not competing solutions but can be combined for the near-future implementations.

\subsection{Development of Cooling Design}

Temperature constrains both the computational performances of xPUs and the efficiency and reliability of power electronics components. As AI workloads scale, aggregate heat dissipation is growing to a level that can no longer be neglected, pushing xPUs towards their thermal limits more rapidly. Cooling thereby becomes a binding systemic constraint on the computational performances of the xPUs that can be sustained.

Chip/Rack-level design typically concerns the thermal fluids across components, where air cooling is being displaced by chip-level liquid cooling to sustain the rack power density beyond what air can dissipate \cite{nvidia2026cooling}. Apart from the on-board cooling loops and heat sink placements \cite{azarifar2024scheme}, this limit is being further pushed by advances at the coolants/heat sink materials \cite{yan2012graphene, qian2021thermcond, dou2025therminterface, wu2025liquidmetal} and thermal channel design \cite{vanerp2020microfluid, kochu2022microfluid, li2025microchannel, wu2025microchannel, sahu2025onchipcooling}, as well as an in-chip microfluidic approach as reported by Microsoft \cite{microsoft2025cooling}.

\begin{figure}[t]
    \centering
    \includegraphics[width=\linewidth]{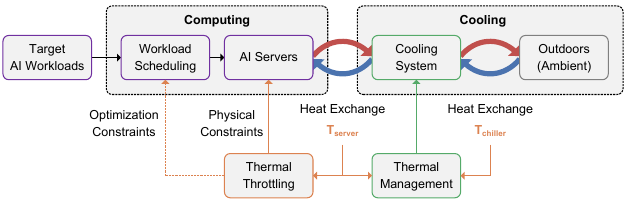}
    \caption{Coupled cooling loops and AI server computing, calling for coordinated electrical-thermal modeling, design and management. Here cooling system involves coolant distribution units (CDUs), chillers and cooling towers.}
    \label{fig_thermal_loop}
\end{figure}

At the system level, the cooling architecture comprises multi-scale heat rejection stages and active cooling controllers, forming coupled \textit{power-temperature cooling loops} through in-rack coolant distribution units (CDUs), chillers and cooling towers, ultimately pushing the heat to the ambient \cite{schneider2025architecture}, as illustrated in Fig.~\ref{fig_thermal_loop}. With the mismatch between the fast and increasingly significant fluctuation of AI workloads and the much slower cooling loops \cite{wang2026thermalmanagement}, the resulting \textit{delay} in heat transport may affect the heat-removal efficiency, in turn constraining the computational performances of xPUs and thus their workload management capability---especially when job schedulers operate independently of facility cooling controllers. Thus, this calls for coordinated electrical-thermal modeling, design and management to ensure the most responsive and resilient system operation.

It is also worth noting that cooling is becoming an interdisciplinary concern more than merely thermal management, given that cooling is itself a substantial and increasingly flexible load on the grid---which can consume as much as 10-20\% of the system power consumption in total \cite{Li2024unseen}. Liquid cooling is effectively enabling the aggregation of the "waste" heat generated by data centers \cite{alissa2025liquid}, while apart from the reuse of it \cite{schneider2026heat, yuan2025heatreuse}, active cooling combined with scheduling of computing workloads is drawing a non-trivial yet comparatively controllable share of the facility power, positioning it as a candidate flexibility resource for grid-interfacing DR and complementing the workload-storage energy balancing strategies \cite{rmi2025flexibility}. To this end, cooling design in AI data centers is a co-design variable alongside voltage class and converter placement, rather than simply a downstream consequence of them.

\section{Stability of Data Center Systems}

The power-delivery paths of modern data centers comprise multiple cascaded power electronic stages, including grid-interfacing rectifiers or uninterruptible power supplies (UPSs), facility-level AC/DC distribution interfaces, server power supply units (PSUs), and downstream point-of-load DC-DC converters \cite{Sun2022DynamicModel}. These stages are regulated by closed-loop controllers and interact through electrical networks. Instability may originate within the data center even when it is connected to a strong utility grid, while rapid workload variations and internal oscillations can propagate through interface converters and affect the external power system \cite{Zhao2026PowerDelivery}. Data center stability is therefore a multiscale interaction problem involving the computing load, cascaded converters, internal distribution system, and utility grid, rather than the isolated stability of any individual device. 

\begin{table*}[b]
    \centering
    \caption{Some Reported Data-Center-Related Oscillation and Load-Reduction Events}
    \label{tab:Instability_events}
    \vspace{-6pt}
    
    \footnotesize
    \renewcommand{\arraystretch}{1.2}
    \renewcommand{\tabularxcolumn}[1]{m{#1}}
    \setlength{\tabcolsep}{2.5pt}
    
    \begin{tabularx}{\linewidth}{
    @{}
    C{1.00}
    C{0.45}
    C{1.35}
    C{1.20}
    @{}
    }
    
    \doubletoprule
    
    \textbf{Events}
    &
    \textbf{Date}
    &
    \textbf{Key Dynamics / Grid Implications}
    &
    \textbf{Cause / Category}
    \\
    
    \midrule
    
    Data center oscillation in Dominion Energy’s power system \cite{Mishra2025DataCenterOscillation}
    &
    Jun. 2022
    &
    Instability of a 10–11~Hz UPS control mode, evolving into a 14.7~Hz sustained oscillation
    &
    UPS--grid control interaction (small-signal instability) with periodic voltage-sag triggers (related to forced oscillation).
    \\
    
    \midrule
    
    ERCOT's large electrical load oscillation in Texas, U.S \cite{Fan2026RealWorld23Hz}
    &
    Oct. 2024
    &
    23~Hz oscillation with a peak-to-peak power of up to 50~MW 
    &
    A poorly damped load-dependent mode associated with the PSU PFC/DC-link voltage control and legacy firmware (small-signal instability).
    \\
    
    \midrule
    
    Oscillation in Meta's data centers \cite{Sun2022StabilityPartI}
    &
    2017
    &
    11~Hz oscillation in data center facility, with 49/71~Hz spectral components 
    &
    Impedance resonance between aggregated server PSUs and the facility distribution network, driven by negative damping from the DC-link voltage-control dynamics (small-signal instability).
    \\
    
    \midrule
    
    Generator--UPS recovery failure at Carrier-1/Prime Data centers in Dallas, U.S. \cite{TierNet2024DallasOutage}
    &
    May, 2024
    &
    Following a generator trip, the UPS switched to off-grid mode and ran on battery power until depletion, resulting in a load loss.
    
    &
    
    (b) Facility-level large-disturbance recovery failure (large-signal instability).
    \\
    
    \midrule
    
    Forced Ringdown in the Midwest, U.S. \cite{NERC2025CharacteristicsLargeLoads}
    &
    2023
    &
    1~Hz active-power perturbations repeatedly excited an 11~Hz grid mode 
    &
    Unintended one-second cycling of data-center power electronics acted as external forcing for the 11~Hz natural grid mode (forced oscillation).
    
    \\
    
    \midrule
    
    Data center load reduction in
    Eastern Interconnection, U.S. \cite{NERC2025LargeLoadLoss}
    &
    Jul, 2024
    &
    Six voltage dips occurred in 82 s, reaching 0.25–0.40~p.u.; approximately 1,500~MW of load was reduced, increasing grid frequency and voltage.
    
    &
    
    A permanent 230~kV fault and repeated auto-reclosing triggered transfer of data-center loads to backup power (grid-interactive impacts).
    \\
    
    \midrule
    
    Data center load reduction in
    Dublin, Ireland \cite{EirGridSONI2025LargeDemandFRT}
    &
    May, 2025
    &
    Data center demand fell by 387~MW, and frequency rose from approximately 50.02 to 50.341~Hz, with RoCoF reaching about +0.26~Hz/s.
    &
    
    A North Wall–Poolbeg 220~kV cable fault triggered data center protection and supply-transfer responses (grid-interactive impacts).
    \\
    
    \doublebottomrule
    \multicolumn{4}{l}{}\\ [-9pt]
        \multicolumn{4}{>{\raggedright\arraybackslash}p{0.99\linewidth}}{\textbf{Note}: UPS---Uninterruptible Power Supply; PSU---Power Supply Units; PFC---Power-Factor Correction;}
    \end{tabularx}
\end{table*}

As summarized in Table~\ref{tab:Instability_events}, reported real-world instability events include the 14.7~Hz oscillation at the Dominion data center \cite{Mishra2025DataCenterOscillation}, the event of approximately 23~Hz observed in the Electric Reliability Council of Texas (ERCOT) system \cite{Fan2026RealWorld23Hz}, and the oscillation around 11~Hz at the Meta data center \cite{Sun2022StabilityPartI}. A data center in the Midwestern United States also experienced periodic 1~Hz disturbances, repeatedly triggering an inherent pattern of approximately 11~Hz \cite{NERC2025CharacteristicsLargeLoads}. Such events can propagate beyond the facility, increase equipment stress, and cause unintended protection operation. The incident in Dallas illustrates an operational recovery failure of data centers after severe power disturbance, where load loss occurred due to the UPS's battery depletion without prompt restoration \cite{TierNet2024DallasOutage}. Disturbance response also presents grid-related stability concerns. A 1500~MW data-center load reduction in the U.S. Eastern Interconnection increased the system frequency and voltage \cite{NERC2025LargeLoadLoss}, while a 387~MW reduction in Ireland raised frequency to 50.341~Hz with a Rate-of-Change-of-Frequency (RoCoF) of approximately +0.26~Hz/s \cite{EirGridSONI2025LargeDemandFRT}. 
These events highlight the need to classify oscillations and disturbance-response failures according to their underlying mechanisms, providing a basis for selecting suitable models, assessment methods, and mitigation measures. To this end, this section first classifies the main types of instability and related disturbance responses, then establishes a stability-oriented three-level framework, in which the dominant mechanisms, modeling and assessment methods, and enhancement strategies are systematically reviewed.

\subsection{Stability Scope and Classification}

\subsubsection{Scope of Instability Issues}
In this article, data center stability is defined as the ability of the interconnected data center electrical system and its supporting grid to maintain bounded electrical variables and recover to an admissible equilibrium or operating trajectory following a disturbance. It covers both internal stability, referring to the dynamic integrity of the power-delivery system inside the facility, and grid-coupled stability, referring to the mutual influence between the data center and the external grid.

\begin{figure}[t]
    \centering
    \includegraphics[width=\linewidth]{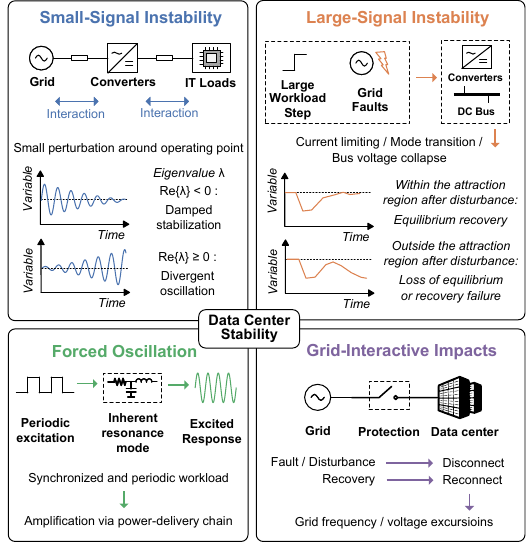}
    \caption{Main instability issues in data centers towards the grid: small- and large-signal stability, forced oscillation, and grid-interactive impacts.}
    \label{Sta_issues}
\end{figure}

The stability-related issues considered here can be grouped into four broad categories, as shown in Fig.~\ref{Sta_issues}. First, small-signal instability includes poorly/negatively damped oscillations around a specific steady-state operating point, caused by interactions among power electronic devices or interactions between devices and the external grid. Such interactions have produced both low-frequency and kilohertz-range resonances in operating data centers \cite{Sun2022StabilityPartI}. Second, large-signal or transient instability concerns the response to disturbance events such as abrupt workload changes, grid or DC bus faults, and UPS mode transitions. These events can cause failure to restore the pre-disturbance operating state, resulting in DC-bus voltage collapse, sustained oscillations, or even widespread load loss \cite{Ross2025EMTModeling}. Third, we include forced oscillations, where synchronized computing activity or other periodic disturbances may excite inherent converters or grid modes, even though they might be stable in the small-signal sense \cite{Ko2026WideAreaOscillation}. Such oscillations can be selectively amplified by the internal power-delivery chain when the excitation frequency approaches the weakly damped mode. The field-observed 14.7-Hz event associated with a data center UPS illustrates that external excitation, a weakly damped internal mode, controller instability, and nonlinear limiting can coexist in the same physical event \cite{Mishra2025DataCenterOscillation}.

In addition, grid-side stability impacts may arise from aggregate transitions in data center power consumption. Severe grid disturbances can trigger protection actions, mode transfers, or large-scale load disconnection of data centers, causing abrupt changes in demand that induce frequency excursions, voltage variations, and power-flow redistribution \cite{NERC2025LargeLoadLoss}. In contrast, poorly coordinated reconnection or load restoration can create secondary demand steps, leading to repeated tripping \cite{ENTSOE2026DataCentres}. Thus, the grid-interactive impacts, like disconnection, reconnection, and protection-induced load changes, are treated as a distinct fourth class of data center stability issues.

Corresponding to Fig.~\ref{Sta_issues}, the reported instability events are also explicitly related to their dominant categories in Table~\ref{tab:Instability_events}. The Meta and ERCOT events are classified under small-signal stability because they are attributed to converter-control and network interactions \cite{Sun2022StabilityPartI, Fan2026RealWorld23Hz}. The Dominion event is also classified primarily as small-signal instability, where periodic external excitation is involved, but the sustained oscillation is attributed to modal instability rather than a purely forced response \cite{Mishra2025DataCenterOscillation}. The Dallas event can be regarded as a large-signal instability due to large-disturbance recovery failure, which normally has no prescribed oscillation frequency \cite {TierNet2024DallasOutage}. The event that occurred in the Midwest U.S. is a typical forced oscillation response \cite{NERC2025CharacteristicsLargeLoads}, while the Eastern Interconnection and Irish load-reduction events are representative cases of grid-interactive impacts, where grid frequency fluctuations are caused by sudden drops in data center load \cite{NERC2025LargeLoadLoss, EirGridSONI2025LargeDemandFRT}.

Note that the categories in Fig.~\ref{Sta_issues} describe mechanisms and disturbance responses of different instability issues in data centers, rather than exclusive frequency bands. Although several reported oscillations in Table~\ref{tab:Instability_events} occur at the sub-synchronous frequencies, the oscillation phenomena in data centers are not limited to the low-frequency range. For example, reference \cite{Sun2022StabilityPartI} reports 11~Hz oscillations observed in the Meta data center, as well as resonance phenomena in the 5–10~kHz frequency range observed at other data centers. Similarly, the forced oscillation response depends on the excitation spectrum and the system’s dynamic characteristics, rather than being limited to a specific frequency band.

\subsubsection{Stability-Oriented Three-Level Framework}
\begin{figure*}[t]
    \centering
    \includegraphics[width=1.7\columnwidth]{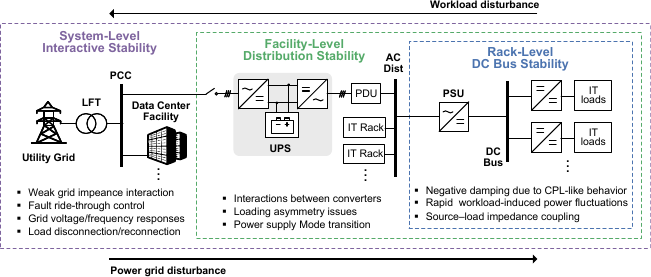}
    \caption{Stability-oriented three-level structure framework of data centers, segmented in terms of rack-level, facility-level, and system-level stability. PCC: Point of Common Coupling; UPS: Uninterruptible Power Supply; PDU: Power Distribution Units; PSU: Power Supply Units; CPL: Constant-power-load.}
    \label{Sta_three_tier}
\end{figure*}

Data center stability encompass dynamics from chip-level voltage regulation to bulk-grid electromechanical responses \cite{Li2025AILoadDynamics, Gyang2026DynamicModel}. A single model that retains all components and time scales is neither computationally practical nor necessary. We therefore organize the main electrical structure of data centers into three analytical tiers, i.e. rack, facility, and system levels, according to the system boundary and the interactions that must be explicitly retained, as shown in Fig.~\ref{Sta_three_tier}. 

At the rack level, the upstream rack supply is normally treated as a stiff or ideal source, whereas the rack DC bus, power shelves or equivalent source converters, downstream point-of-load (PoL) converters, and IT loads should be explicitly represented \cite{Lin2024HVDCServerRacks}. The key concern is whether the local DC bus remains stable under the source–load impedance coupling, the constant-power-load (CPL)-like behavior, and rapid workload-induced power fluctuations \cite{Sun2022StabilityPartI, Li2025AILoadDynamics}. 

At the facility level, the internal distribution network is further involved, so the study boundary might cover multiple rack-level PSUs, energy-storage interfaces, electrically relevant cooling loads, and grid-side converter devices (UPSs or SSTs) \cite{Gyang2026GridIntegratedReview}. This level focuses attention on converter–converter and converter–network interactions inside the facility, thus capturing phenomena that cannot be inferred from a single rack, such as interactions between UPSs and PSUs, loading asymmetry and power supply transition among normal, double-conversion, bypass, and backup modes \cite{Sun2022StabilityPartII}.

At the system level, the external power system is included together with the data center at the PCC. Modeling of data center internal facilities should reproduce their terminal active/reactive power responses, input impedance, ride-through characteristics, and protective disconnection/reconnection logic, thus being sufficient to support studies on the impact of weak grid interactions, grid mode excitation, fault ride-through control, voltage and frequency responses, and coordinated load switching \cite{ENTSOE2026DataCentres}. 

These three tiers are bidirectionally coupled. Workload variations propagate outward from racks to the grid, whereas grid disturbances propagate backward into the data centers \cite{Zhao2026PowerDelivery, JimenezRuiz2025TransientModel}. The appropriate tier is determined according to the disturbance origin, dominant interaction, and the variables used to judge stability. Hence, a component can be modeled explicitly in one tier and represented by an equivalent in another. This hierarchical view provides the basis for the stability mechanisms, model fidelity, and assessment methods which will be discussed later. 

Note that the three-level framework represents stability-study boundaries rather than fixed physical architectures, so it is applicable to both conventional AC-distributed and emerging DC-distributed data centers, although the boundaries between levels may differ. In conventional AC architectures, PSU stages provide a relatively clear dynamic interface between the facility-level distribution network and rack-level DC buses \cite{Sun2022StabilityPartI}. In DC-distributed architectures, rack buses are connected more directly to a common facility DC network, thereby weakening this separation \cite{kolar2026sst}. As a result, rack- and facility-level stability analyses may partially overlap, although the latter must still retain distribution impedance, multiple racks, and upstream grid-side converters.

\subsection{Instability Mechanisms}

\subsubsection{Rack-Level DC-Bus Stability}
Rack-level instability is primarily governed by the dynamic interaction between the regulated IT loads, the finite energy storage, and the control bandwidth of the rack DC power supply, as summarized in Table~\ref{Instability_mechanism}. Although the upstream source might appear stiff in steady state, its closed-loop output impedance, together with bus capacitance, cable inductance, and downstream PoL converter dynamics, determines whether disturbances are damped or amplified \cite{Lin2024HVDCServerRacks}.

A key mechanism is the constant-power-like behavior of IT loads \cite{Kwasinski2011InstantaneousCPL}. Within the effective control bandwidth of downstream converters, the load input current $\mathrm{i}$ approximately follows $\mathrm{i = P/v}$, where $\mathrm{P}$ is the constant input power consumed by the IT load, $\mathrm{v}$ is the DC-bus voltage, and small-signal linearization near the steady-state operating point $\mathrm{(V, I)}$ can yield:
\begin{equation}
\label{CPL}
\mathrm{\Delta i = - \frac{P}{V^2}\Delta v}
\end{equation}

Thus, a reduction in bus voltage causes the load to draw more current to maintain its output power, increasing the voltage drop across the source impedance and further depressing the bus voltage. This negative incremental impedance reduces damping and further destabilizes the system. However, the IT load should not be regarded as an ideal CPL over the entire frequency range. Above the regulating bandwidth, the terminal behavior depends on the PoL converter's internal dynamics \cite{Li2025AILoadDynamics}. Note that parallel operation of source- or load-side converters can introduce additional coupling through the common DC bus. Voltage regulation, current sharing, and droop control may create weakly damped modes, even when each converter is stable in isolation \cite{Zhao2025MultiConverterInteraction}. 

For a rack-level system that is locally stable in the small-signal sense, its large-signal recovery presents different results. Under large workload steps or severe voltage disturbances, the DC-link capacitor must temporarily supply the power imbalance. If the source converter reaches its current limit and cannot supplement the stored energy rapidly, the bus may experience deep voltage excursions or even collapse \cite{Li2025AILoadDynamics, Gui2021LargeSignal}. 

AI workloads make these mechanisms increasingly relevant because their electrical demand is both rapid and highly synchronized \cite{Choukse2025PowerStabilization}. Slow workload changes mainly shift the operating point and hence the incremental load impedance and stability margin. Abrupt workload transitions can initiate large DC-bus excursions, whereas periodic or synchronized computing activity can act as a sustained excitation of weakly damped rack modes, leading to forced oscillations \cite{Valverde2025AIForcedOscillation}. When many racks follow correlated power profiles, the resulting fluctuations will no longer remain local, but rather propagate through the PSUs and become an input disturbance to the facility distribution system, linking the facility-level issues discussed next.

\subsubsection{Facility-Level Distribution Stability} As shown in Table~\ref{Instability_mechanism}, facility-level instability is mainly caused by the interaction of multiple regulated converters through a shared distribution network. A central mechanism is impedance interaction. The upstream UPS or front-end converter does not behave as an ideal voltage source, while downstream PSUs may exhibit negative or non-passive impedance characteristics. When the magnitude and phase of the source impedance and load impedance become unfavorable near a network or filter resonance, oscillation might occur \cite{Sun2022StabilityPartI}. Low-frequency modes are often associated with DC-voltage regulation or constant-power behavior, whereas higher-frequency resonances arise from input filters or digital delays \cite{Sun2022StabilityPartII}. Large disturbances introduce further nonlinear mechanisms, where grid-voltage dips, rapid workload changes, and UPS mode transfers disturb the power balance between upstream sources and downstream loads, leading to deep bus-voltage excursions and transient oscillations \cite{Sun2022DynamicModel, Gyang2026DynamicModel}.

Normally, a large number of racks are connected in parallel to the distribution network, but aggregating all racks into a single equivalent load might hide internal modes \cite{Zhu2023PFCAggregation}. A current disturbance in one branch produces a voltage variation on the common bus, which is then sensed by other controlled converters and modifies their power demand. This feedback can create common-mode oscillations involving many racks or more localized differential modes between neighboring UPS–PSU groups \cite{Sun2022StabilityPartII}. Moreover, in conventional AC-distributed facilities, additional coupling may arise from single-phase PSUs connected across a three-phase network. Uneven phase loading and zero-sequence paths can reduce stability margins and produce dynamics that are not captured by balanced positive-sequence models \cite{Sun2022StabilityPartI}. In DC-distributed architectures, these effects are reduced, but the direct connection of multiple racks to a common DC network strengthens the cross-rack interactions, extending the local rack concerns to the facility-wide DC microgrid stability issues \cite{Lin2024HVDCServerRacks, Zhu2024SystemLevelReduction}.

\subsubsection{System-Level Interactive Stability}
At the system level, the dominant issue is the bidirectional dynamic coupling between the data-center facility and the external grid, which is also stated in Table~\ref{Instability_mechanism}. One key mechanism is the interaction between the grid impedance and the data center’s aggregated input characteristics. The control dynamics of interface converters affect the inherent grid modes associated with transmission lines, synchronous generators, or renewable energy sources, especially under weak-grid conditions \cite{VasquezPlaza2026WeakGrid, Arifin2026GridIntegrated}. Thus, an internally stable data center can also excite or amplify an external grid mode, while grid dynamics can in turn destabilize its front-end converters. Periodic workload variations provide another coupling path. Synchronized computing activity can propagate through the internal power-delivery chain and appear at the PCC. If the forcing frequency is close to a weakly damped facility or grid mode, the disturbance will be selectively amplified \cite{Zhao2026PowerDelivery, Ko2026WideAreaOscillation}.

During large grid disturbances, such as voltage dips, phase jumps, or frequency events, the grid-interfacing converter may reach its current limit \cite{Ross2025EMTModeling}, while downstream loads continue to demand nearly constant power. The resulting mismatch must be supplied by DC-link capacitors, UPS batteries, or other storage devices \cite{Ross2026GFMESS}. Insufficient stored energy and inappropriate mode transitions will cause DC-voltage excursions and eventual fault-ride-through failures. Furthermore, such a severe disturbance event can cause data centers to trip within a short period, producing an abrupt reduction in demand and consequent frequency/voltage excursions \cite{NERC2025LargeLoadLoss}. Furthermore, rapid and synchronized load restoration might trigger voltage depression, frequency reduction, and even repeated tripping \cite{JimenezRuiz2025TransientModel}. System-level stability is therefore governed by a closed disturbance chain in which grid conditions determine the operational responses of data centers, while the resulting power transition modifies the external grid.

\begin{table*}[t]
    \centering
    \caption{Instability Mechanisms in the Three-Level Framework of Data Centers}
    \label{Instability_mechanism}
    \vspace{-6pt}
    
    \footnotesize
    \renewcommand{\arraystretch}{1.2}
    \renewcommand{\tabularxcolumn}[1]{m{#1}}
    \setlength{\tabcolsep}{3pt}
    
    \begin{tabularx}{\linewidth}{
    @{}
    C{0.55}
    C{1.05}
    C{1.45}
    C{0.95}
    C{1.00}
    @{}
    }
    \doubletoprule
    
    \textbf{System Level}
    &
    \textbf{Main dynamic interface}
    &
    \textbf{Dominant destabilizing mechanisms}
    &
    \textbf{Typical conditions}
    &
    \textbf{Observable consequence}
    \\
    
    \midrule
    
    Rack level
    &
    DC bus between PSUs and IT loads
    &
    Negative incremental impedance;\linebreak
    Insufficient DC-link energy buffering;\linebreak
    Interaction among parallel converters
    &
    Large workload step; Source-voltage variation;\linebreak
    Control bandwidth overlap
    &
    DC-bus oscillation;\linebreak
    Deep voltage sag or even collapse;\linebreak
    Local protection action
    \\
    
    \midrule
    
    Facility level
    &
    Shared AC or DC distribution network
    &
    Impedance interaction;\linebreak
    Unbalanced load and zero-sequence paths;\linebreak
    Multi-rack aggregation
    &
    Uncoordinated workload;\linebreak
    Weak internal bus;\linebreak
    UPS mode transition
    &
    Multi-converter resonance;\linebreak
    Distribution bus oscillation;\linebreak
    Uneven power sharing
    
    \\
    
    \midrule
    
    System level
    &
    PCC between the interface converter and the external grid
    &
    Interaction between aggregated data-center and grid dynamics;\linebreak
    Workload-induced inherent grid oscillation;\linebreak
    Disturbance-driven load transfer
    &
    Weak-grid operation;\linebreak
    Periodic workload excitation;\linebreak
    Fault ride-through failure
    &
    Oscillation at PCC;\linebreak
    Abrupt load loss;\linebreak
    Voltage/frequency excursion 
    \\
    
    \doublebottomrule
    \end{tabularx}
\end{table*}

\subsection{Stability Modeling and Assessment Methods}

\begin{figure}[t]
    \centering
    \includegraphics[width=\linewidth]{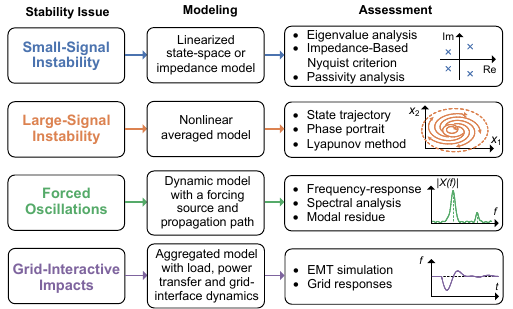}
    \caption{Modeling and assessment methods for different data center instability issues.}
    \label{Sta_model_assess}
\end{figure}

The wide range of electrical architectures, time scales, and operating modes in data centers makes it difficult to define a universally applicable stability model. Model fidelity should be selected according to the stability phenomenon and the time scale of interest. This section first reviews the hierarchy of models used to represent data-center dynamics and then relates these models to the corresponding stability-assessment methods, as given in Fig.~\ref{Sta_model_assess}.

\subsubsection{Stability Modeling}
Stability-oriented data-center modeling can begin with the representation of computational workload and electrical load, which describe two related but distinct processes. The workload model specifies how the computing activities create a time-varying power request, whereas the electrical model determines how that request appears at the terminals of different levels. Workload demand can be represented by real-world measured data series, or by stepwise/pulsed/periodic profiles, depending on the study objectives (operating-point changes/large transients/sustained excitation) \cite{JimenezRuiz2025TransientModel}. For example, a recent system-level study has used aggregated AI workload traces as explicit excitation signals for dynamic analysis \cite{Zhao2026PowerDelivery}. 

The electrical load can be represented at several levels of detail. An ideal CPL provides a concise description of tightly regulated downstream converters and is particularly useful for exposing negative incremental impedance. However, its validity is limited to the frequency band over which the converter can maintain its power demand. More realistic reduced models introduce further considerations, such as finite response times, voltage-dependent behavior, current limits, etc \cite{Weber2025PERC1}. When the internal control loops materially influence the oscillation mode of interest, the IT load must instead be represented by an averaged converter model, since such a model preserves the key control dynamics without resolving every switching action \cite{Sun2022DynamicModel}. After linearization, the averaged models support time-domain state-space and frequency-domain terminal-impedance formulations for small-signal stability analysis. It is worth mentioning that the appropriate frequency-domain models also depend on the electrical structure, where conventional $\mathrm{dq}$-domain models are effective for balanced three-phase converters because they transform the fundamental-frequency periodic steady state into a time-invariant operating point \cite{Sun2009SmallSignalMethods}, but sequence-domain formulations become preferable when positive-, negative-, and zero-sequence channels must be separated and assessed explicitly \cite{Sun2022StabilityPartII}. For periodically time-varying systems, advanced harmonic linearization models are required \cite{Zhang2019HarmonicLinearization}. These distinctions are important in data centers because conventional AC architectures combine three-phase UPS equipment with large numbers of single-phase PSUs, whereas future DC architectures shift the dominant coupling toward common DC buses and cascaded DC–DC stages. Thus, the model should follow the relevant port structure rather than apply one coordinate system uniformly across all levels.

Large-signal stability studies require models that retain the nonlinear power or energy relationships beyond a single operating point. A common choice is the nonlinear averaged converter model, obtained by averaging the switching behavior but preserving the nonlinear dependence of the system, thus being able to describe the dynamics responsible for equilibrium recovery after large disturbances \cite{Gui2021LargeSignal}. Note that when current limiting or operating-mode transitions are involved, the model must be extended into a piecewise or hybrid representation with mode-dependent equations and transition logic \cite{Ross2025EMTModeling}. These nonlinear models provide the basis for the phase-portrait-based state trajectory and Lyapunov-function-based stability analysis as discussed later. 

Forced-oscillation studies require explicit representation of both the forcing source and the transmission pathway. The excitation source primarily arises from changes in the AI workload, where ideal sinusoidal and square-wave models are easier for resonance analysis, but stochastic periodic workload models better capture cycle-to-cycle variability and the aggregation of multiple AI tasks \cite{Ko2026WideAreaOscillation}. The resulting system responses can be assessed through modal transfer functions to reveal the sensitivity of a given rack to a specific mode \cite{Valverde2025AIForcedOscillation}, or through converter-oriented admittance models to involve converter control dynamics \cite{Chen2025GFMGFLForcedOscillation}.

As for grid-interactive impact, it requires an aggregated PCC-oriented model that captures both continuous power dynamics and discrete operating transitions during grid disturbances \cite{Weber2025PERC1}. The event-driven logic should also be elucidated, which governs transitions among normal operation, ride-through, disconnection, and restoration. 

Detailed EMT or switching models remain necessary when the study concerns high-frequency resonances, switching harmonics, or fast faults. However, their computational burden restricts the number of devices and the simulation duration \cite{Ross2025EMTModeling}. A supraharmonic study illustrates the extreme end of this hierarchy \cite{Afzaal2025SupraharmonicEMT}, where resolving emissions in the kilohertz range requires explicit switching devices, high-frequency filters, and sufficiently small simulation time steps, whereas such detail would be unnecessary for electromechanical frequency studies. In contrast, for large-scale phasor-domain dynamic studies where the concerned dynamics evolve much more slowly than converter switching, reduced-order positive-sequence models are appropriate. They enable large transmission networks and multiple data centers to be simulated efficiently, but their validity must be checked against a detailed EMT model \cite{Zhao2026PowerDelivery}. 

Finally, measurement-based data-driven models provide an alternative when proprietary converter parameters are unavailable. Terminal impedance measurement, transfer-function fitting, and system identification can characterize the data center as a black-box dynamic subsystem \cite{Sun2022StabilityPartI, Viththarachchige2026LoadModelingReview}, while modal-identification techniques can extract oscillatory components from operational data \cite{Mishra2025DataCenterOscillation}. These approaches are valuable for model calibration and event diagnosis, but their conclusions remain dependent on the operating conditions represented in the data. Therefore, they are best regarded as complements to, rather than replacements for, physics-based models.

\subsubsection{Stability Assessment Methods}
For small-signal stability, linearized state-space eigenvalue analysis provides direct information on modal frequency, damping, and participation factors. Eigenvalue loci can also be obtained to determine how the stability margin changes with operating conditions \cite{Zhao2026PowerDelivery, Anand2013ReducedOrder}. However, the physical origin of a mode may become difficult to interpret, and model order increases sharply in a large system containing many racks and converter stages. Impedance-based analysis offers a more modular alternative \cite{Sun2022StabilityPartI}. The system is divided at a selected electrical interface, and stability is inferred from the impedance ratio between the two sides by the Nyquist criterion or impedance magnitude/phase relationships in Bode plots \cite{Lin2024HVDCServerRacks}. This approach is particularly suitable for source–load coupling and cascaded conversion stages since each subsystem can be modeled or measured separately. A scalar impedance ratio may be sufficient for a DC bus or a single-phase channel, while three-phase and sequence-coupled systems generally require matrix-based formulations and the generalized Nyquist criterion \cite{Sun2022StabilityPartII}. In addition, passivity-based assessment can further provide a conservative, interface-oriented assessment of stability when the exact grid impedance is uncertain \cite{Meng2026Dissipativity}. However, all these methods remain local to the operating point used for linearization and cannot assess whether the system will recover after a sufficiently large disturbance.

For large-signal stability, the key question is whether the post-disturbance operating trajectory remains within a recoverable region. Based on the nonlinear averaged model, phase-portrait-based state trajectory methods can distinguish convergence toward the desired equilibrium from movement toward collapse, and can reveal how the critical boundary changes with load power or stored energy \cite{Xu2025DABStateTrajectory}. Lyapunov direct methods provide a more formal interpretation of the same problem. Rather than following one disturbance trajectory, they seek a scalar measure that decreases along all trajectories within a specified region. This enables estimation of an attraction region as a sufficient stability boundary, although constructing a useful Lyapunov function becomes difficult for high-order systems \cite{Herrera2017DCMicrogrid}. Thus, in practical data-center studies, nonlinear analytical methods are most useful for reduced rack- or converter-level models, while detailed EMT simulation is often needed to verify the resulting conclusions in the complete facility \cite{Viththarachchige2026LoadModelingReview}.

The assessment of forced oscillations requires separating the excitation source from the dynamic path that amplifies it. Eigenvalues indicate the natural modes of the system, but do not by themselves show whether a particular workload signal can excite those modes. By input–output frequency response analysis from workload power to bus voltage, UPS current, or PCC power, we can identify frequencies at which the internal power-delivery chain acts as an amplifier \cite{Valverde2025AIForcedOscillation}. When measured waveforms are available, spectral analysis can identify dominant frequency components, while Prony analysis or dynamic mode decomposition (DMD) helps to estimate mode damping \cite{Ko2026WideAreaOscillation, Mishra2025DataCenterOscillation, MR-DMD}. However, measurement alone cannot distinguish a forced oscillation from small-signal oscillation. Hence, forced-oscillation diagnosis is more reliable when frequency-domain propagation analysis is combined with measurement-based modal identification.

For grid-side stability impacts, a grid fault may cause the data center facility to remain connected, transfer part of its demand to local storage, or totally disconnect from the grid \cite{Ross2025EMTModeling}. These different responses produce different active- and reactive-power trajectories at the PCC and different effects on the external network. Dynamic network simulations are used to evaluate the resulting frequency, voltage, and power-flow responses \cite{Arifin2026GridIntegrated}. For instance, sudden disconnection appears to the grid as a loss of demand and may raise system frequency and local voltage, whereas rapid reconnection introduces a new load step that can depress both. The relevant measured metrics depend on the event under study, including frequency extrema, rate of change of frequency, voltage recovery time, and the magnitude and duration of power-flow redistribution \cite{JimenezRuiz2025TransientModel, Gyang2026DynamicModel}.

In practice, no single assessment method can cover the full range of data center dynamics. Time-domain simulation results present whether a stability problem occurs, while analytical methods are then needed to explain why it occurs. A credible assessment framework therefore combines event reproduction with a method capable of tracing the observed behavior back to its dominant physical mechanism.

\subsection{Stability Enhancement Strategies}
\begin{figure*}[t]
    \centering
    \includegraphics[width=\linewidth]{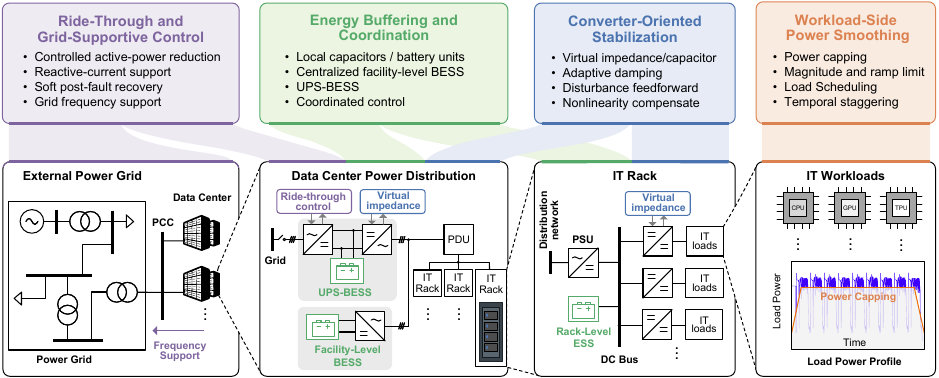}
    \caption{Stability enhancement strategies for data centers — Detailed implementation measures and typical application scenarios in different system tiers. PCC: Point of Common Coupling; UPS: Uninterruptible Power Supply; BESS: Battery Energy Storage System; PDU: Power Distribution Units; PSU: Power Supply Units;}
    \label{Sta_Enhance}
\end{figure*}

Stability enhancement in data centers can intervene at different stages of the disturbance path. The appropriate strategy depends on the dominant instability mechanisms. Accordingly, the following discussion organizes the available strategies from workload-side mitigation to grid-interfacing converter control. Their detailed implementation measures and typical application examples are shown in Fig.~\ref{Sta_Enhance}.

\subsubsection{Workload-Side Power Smoothing}
Workload-side power smoothing seeks to reduce the amplitude and coherence of the disturbance at its computational origin. Rather than allowing server demand to vary freely, computing resources are operated to lower the magnitude and ramp rate of workload transitions, so that the aggregate power profile changes more smoothly. This can mitigate the instantaneous burden placed on rack DC buses and upstream converters. Specifically, power capping or processor-frequency adjustment can provide a relatively direct reduction in demand, while job scheduling acts by delaying, pausing, or reallocating tasks \cite{Colangelo_2026}. For workloads running across many servers, temporal staggering can reduce the likelihood that correlated computing activity will produce a coherent power oscillation \cite{Choukse2025PowerStabilization}. More advanced schemes achieve a closed loop between computing operation and electrical demand, which use measured facility power, workload status, and external grid requests to determine how much computational flexibility is available without violating service requirements \cite{Gyang2026DynamicModel}. This result indicates that software-level control can provide a meaningful first layer of power smoothing without requiring additional energy-storage hardware.

However, workload control is suitable only for schedulable variations and is constrained by computational latency. It is generally too slow to suppress converter resonances or manage the power imbalance caused by a grid fault. As a result, workload-side smoothing serves only as a supplement to other stabilization measures.

\subsubsection{Converter-Oriented Stabilization}
Converter-oriented stabilization seeks to modify the dynamic behavior presented at an electrical interface of power converters. In small-disturbance conditions, the main objective is to prevent converter controls from supplying negative damping near a network resonance. Active damping achieves this effect through supplementary feedback, emulating the action of a resistor or impedance without introducing the continuous losses associated with physical damping networks \cite{Rahimi2009ActiveDamping}. The method can be implemented at the load converter to counteract the destabilizing negative incremental impedance, or at the source converter to introduce positive damping near a critical resonance frequency \cite{Zhang2015LoadSideVI, Lu2015VirtualImpedance}. 

A constant virtual resistance applied over a broad frequency range may slow voltage regulation and disturb the power sharing, while frequency-selective virtual impedance addresses this conflict by adding damping mainly around the poorly damped mode \cite{Guo2017FrequencyDependentVI}. In data center DC networks, the resonance frequency may also shift as cable length and rack power change, so fixed tuning will lose effectiveness, motivating adaptive schemes that estimate the dominant oscillation frequency to retune the controller online \cite{Su2025AdaptiveVI, Fan2025SOGIFLL}. Moreover, recent approaches aim at separating stabilization from the main regulation task. One data-center-oriented example uses the flying capacitor of a three-level Buck converter as a controllable oscillation-energy buffer \cite{Su2026IRFC}. By redirecting the resonant power into this internal capacitor, the converter input impedance can be reshaped without substantially compromising the output dynamics seen by the chip load. Virtual-capacitor methods follow a related principle, synthesizing additional capacitive behavior at source and load ports so that short-term power imbalance is absorbed locally rather than propagated through the bus \cite{Su2026VirtualCapacitor}.

During large disturbances, the control objective must instead ensure that the converter trajectory remains within a recoverable region, so large-signal stabilization acts directly on the nonlinear power balance. In rack-level systems, load current feedforward can be incorporated into conventional cascaded voltage–current control to provide stability guarantees beyond a single operating point \cite{Gui2021LargeSignal}. More explicitly nonlinear controllers use the converter states to shape the entire post-disturbance trajectory. Backstepping and feedback-linearization methods compensate the CPL nonlinearity, often with an observer to accommodate uncertain load power \cite{Xu2019CompositeNonlinear}. Their practical limitations differ, but the common purpose is a larger region of attraction to prevent the current and DC voltage from crossing the boundary between recovery and collapse.

\subsubsection{Energy Buffering and Coordination}
By coordinating the internal power-conversion stages and energy buffers in data centers, disturbances can be prevented from propagating through the entire facility. This requires a deliberate allocation of response across different time scales and placements. Local capacitors or backup battery units located close to the rack or low-voltage DC bus can absorb the sharpest transient disturbances before they reach upstream PSUs and distribution feeders \cite{Li2025AILoadDynamics}, while centralized UPS batteries and dedicated energy-storage systems can serve a larger group of loads with higher utilization and facilitate power balance over longer intervals \cite{Ross2026GFMESS, ENTSOE2026DataCentres}, but the response is affected by intervening converters and network impedance.

Coordination is also needed between storage and the existing UPS infrastructure. Conventional UPS batteries are reserved primarily for continuity of supply, so they need to respond instantaneously when the upstream source cannot maintain the DC link, but should not be responsible for smoothing sustained deviations that would compromise backup capability \cite{Ross2025EMTModeling}. A further improvement is to coordinate electrical storage with scheduled computing demand, since workload scheduling and battery response serve as different forms of flexibility and need not compete for the same disturbance component. A demonstrated strategy separates facility power fluctuation by time scale, where workload migration or temporal shift reshapes the portion that can be adjusted without violating service requirements, while UPS storage compensates the residual component subject to state-of-charge and cycling constraints \cite{Yang2018HolisticDemandResponse}.

In addition, the same facility-level controller must coordinate recovery after the disturbance. If several UPSs or storage converters recharge simultaneously, the restoration process can create a secondary power step and undo the original smoothing action. The charging power should be ramped and distributed according to the state of charge and the available redundancy \cite{JimenezRuiz2025TransientModel}. Similar coordination is also required when parallel converters enter current limiting or transfer between normal and backup modes, since uncoordinated transitions may produce circulating power or uneven loading.

\subsubsection{Ride-Through and Grid-Supportive Control}
As data centers become systemically significant loads, their grid fault ride-through control must satisfy two objectives simultaneously, i.e., maintaining uninterrupted supply to critical IT loads and avoiding an abrupt loss of demand at the grid interface. Conventional UPS protection is designed mainly around equipment-tolerance curves, and normally transfers the facility to backup supply after relatively shallow voltage depressions \cite{Ross2025EMTModeling}. However, such a response exacerbates a normally cleared grid fault due to sudden removal of substantial load. Thus, emerging requirements favor controlled continuation of grid connection whenever converter and equipment limits permit \cite{CIGRE2026PEILTechnicalFoundations}.

The achievable ride-through control responses depend on the grid-interface device and operating mode. In double-conversion operation of UPS, the front-end rectifier provides a controllable interface through which fault current, active-power draw, and reactive-power exchange can be shaped. In bypass mode, the UPS converter is no longer dynamically interposed between the grid and downstream PSUs, so the facility response depends more directly on load tolerance and transfer logic \cite{Sun2022DynamicModel}. In future DC architectures, a central rectifier or SST feeds a common DC network, while storage connected to that bus can support the load without requiring the entire facility to disconnect from the grid \cite{ocp2026lvdc}.

During severe grid disturbances such as voltage sags, the grid-interfacing converter cannot usually maintain full active power without exceeding its current limit. A practical strategy is to reduce active-power import in a controlled manner, reserve part of the available current for reactive support, and make the UPS battery or another local buffer supply the remaining load demand \cite{Shamseldein2026ThreeMode}. Moreover, under unbalanced faults, sequence-selective control is needed to avoid excessive negative-sequence current and power oscillations \cite{Nasr2023NegativeSequenceControl}. Post-fault recovery is equally important, because an immediate return to the pre-fault demand may create a second voltage or frequency disturbance. The recovery should be staged and coordinated with DC-link restoration and energy storage \cite{ENTSOE2026DataCentres}.

After secure ride-through has been established, the data center facility may further provide deliberate grid support. Under weak-grid conditions, direct power control of the grid-interfacing rectifier can support the PCC voltage via reactive-current control \cite{VasquezPlaza2026WeakGrid}. UPS batteries can rapidly modulate the facility demand to provide fast frequency response \cite{Alapera2019UPSFFR}. Therefore, ride-through and grid support performance should be specified through the desired power trajectory at the PCC, which makes it imperative to establish a specification for data centers in the future.

\section{Outlook}

Building on the discussed grid-to-chip trends in data centers, this chapter outlines the outlooks for technology development: the evolution of computational loads, the grid architecture, how power and information infrastructures are converging, as well as how data centers should be aligned with sustainability from a grid perspective.

\subsection{Evolving Computational Loads}

Computational loads have been evolving steadily, and the grid-to-data-center consolidation remains to be fully exploited. This section discusses the infrastructural impacts of load evolution on the hierarchies, which involve: the scaling challenges associated with computational loads faced by data centers, needs of power supplies for leading-edge and emerging computing, as well as the potentials of grid-supportive and grid-interactive data centers emerging from the flexibility of data centers and equipped with energy storage. 


\subsubsection{Scaling of computational loads}

The trend toward larger data centers is driven by the need to train large-scale AI models and the economies of scale associated with power usage effectiveness (PUE). PUE is a metric that measures the ratio of a data center’s total energy consumption $\mathrm{E}_\mathrm{DC}$ to its IT equipment's energy consumption $\mathrm{E}_\mathrm {IT}$, indicating how efficiently the data center operates as \cite{googlepue}:
\begin{equation}
    \label{eq:pue}
    \mathrm{PUE} = \frac{\mathrm{E}_\mathrm{DC}}{\mathrm{E}_\mathrm{IT}} > 1.0
\end{equation}
where, a PUE of 1.0 is the ideal benchmark, but it is practically unattainable given the intrinsically underlying energy consumption of supportive non-IT facilities.

Large-scale data centers are known to have lower PUE values. According to \cite{masanet2020recalibrating}, between 2010 and 2018, the adoption of hyperscale data centers increased the computing power by 550\%, while global data center power consumption rose by only 6\%.
In an effort to develop more advanced AI models, there is an ever-growing trend toward increasing the number of GPUs used for training. On July 22, 2026, OpenAI announced plans to build a 3.2~GW AI data center in Georgia, called Project Camellia \cite{OpenAI_Camella}.

Furthermore, with the shift toward multimodal AI, which integrates and simultaneously processes and understands not only language but also images, audio, video, code, sensor data, and so on. The number of required parameters and the volume of training data have increased dramatically. The challenge in this direction is the physical constraints of building large-scale data centers and integrating it with power systems. 
Therefore, methods for distributing AI model training resources across multiple data centers have also been proposed. Google used a large fleet of TPUs across multiple data centers when training the Gemini Ultra model \cite{Google_Gemini}.


\subsubsection{Power supply for leading-edge and emerging computing}

Power supply is also becoming a compelling constraint as computing paradigms scale. Quantum computing, for instance, is promising in terms of computational speed, yet its energy footprint remains to be characterized \cite{hsu2015quantum, chen2023quantum}. Though proof-of-concept results have been reported \cite{desdentado2024quantum}, the requirements on power supply still need to be confirmed before applications. It is noteworthy that operational conditions also plays a non-negligible role: up-to-date quantum computers require extremely low temperatures (e.g., below 1~K) and stringent noise/electromagnetic conditions \cite{hsu2015quantum, teo2021quantum, itestquantum}, thereby posing critical challenges in the power supply design. Likewise, further computing paradigms may introduce new electromagnetic, mechanical, or thermal demands, that should be understood from the underlying physics.


\subsubsection{Grid-interactive data centers}

A further research direction is the evolution of data centers from passive loads into grid-interactive facilities. The coordinated use of flexible computing demand, cooling-system thermal inertia, and energy storage could enable a data center to regulate its power at the PCC in a manner similar to a virtual power plant (VPP) \cite{Morovati2025VPPDataCenters, Crozier2025GridFlexibility}. The key challenge is to ensure that such flexibility does not compromise internal service requirements or introduce new stability risks. At the same time, the dense power electronics inside data centers create an opportunity for converter-embedded grid sensing. Power electronic devices are expected to support local measurement and estimation of key metrics like grid strength, oscillatory modes, voltage unbalance, etc, which could be further used for adaptive damping or power-support functions in real time \cite{Mu2025SelfMeasurement, Marini2025RealTimeImpedance}. Realizing this vision will require reliable measurements, cybersecurity safeguards, and standardized rules defining which services a data center can provide without jeopardizing its primary functions.

\subsection{Evolving Grid Architectures for Data Centers}

The evolution of grid architectures also introduces new opportunities. As the grids move towards distributed operation, lower rotational inertia and more varied interactions, future data centers should still conform to both functionality and standardization that are to be reformed. One example is the emerging SAFEr Grid architecture, and another can be following standardization considerations like grid specifications and fault-ride-through (FRT) requirements. From a socio-economic perspective, responsibilities are also unraveled that each stakeholder should fulfill regarding the data center consolidation.


\subsubsection{The SAFEr Grid infrastructure}

As physical inertia of power systems reduces with the rising penetration of distributed energy resources (DERs), the Store-and-Forward Grid (SAFEr Grid) concept proposes compartmentalizing power systems into asynchronous sub-grids (A-grids), which exchange energy among each other through power electronics energy routers and storage rather than being globally synchronized \cite{safergrid2026}. This promises to improve system resilience and opens new socio-economic opportunities with higher potential energy flexibility. Data centers are a close match for this concept: the power supply chain inside them is already power-electronics-based and frequency-agnostic, and their MV-SST boundary with a backup-battery unit (BBU) is ready to function as energy routers. At its core, data centers embody the "third operational mode" of power systems beyond the conventional islanded or grid-connected modes, where A-grids operate "autonomously" from an energy perspective but remain physically interconnected as SAFEr Grid introduces---data centers are desired to isolate the internal power fluctuation to ensure grid stability, and vice versa. In this context, data centers can be positioned as a representative near-ready proving ground for the concept, though the realization requires the internal hierarchy to address both the functionality of data center loads and energy-routing resilience commitments in the society as introduced from system level.

\begin{figure}[t]
    \centering
    \includegraphics[width=\linewidth]{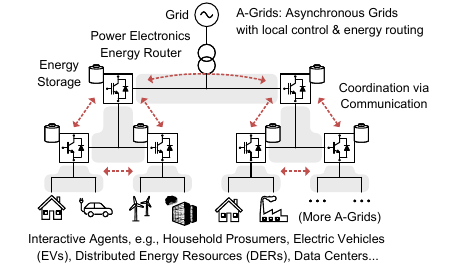}
    \caption{The Store-and-Forward energy routing for future asynchronous power grids (the SAFEr Grid architecture) that aims to maximize the exploitation of energy flexibility in asynchronous sub-grids (A-grids) \cite{safergrid2026}. Data centers inherently align with the concept by virtue of the power supply facilities.}
    \label{fig_safer_grid}
\end{figure}


\subsubsection{Grid specifications of data centers}

For ride-through and grid-supportive requirements of data centers, directly copying generator-side requirements may be inappropriate because a data center remains an uninterruptible load whose internal energy reserve is finite and whose protection was originally designed around continuity of service \cite{NERC2026GapAssessment}. The specification should define the admissible load reduction during a disturbance, the active–reactive current priority, the duration supported by local storage, and the permitted post-fault recovery rate \cite{Shamseldein2026ThreeMode}. It should also recognize differences among centralized UPS systems, bypass-operated AC systems, distributed backup systems, and emerging DC architectures \cite{ocp2026lvdc}. A fit-for-purpose requirement should ensure that the facility remains secure, while preventing its transfer or recovery logic from amplifying the original grid disturbance \cite{NERC2026RiskMitigation}.


\subsubsection{Requirements on FRT capability}

\begin{figure}[t]
    \centering
    \includegraphics[width=0.85\linewidth]{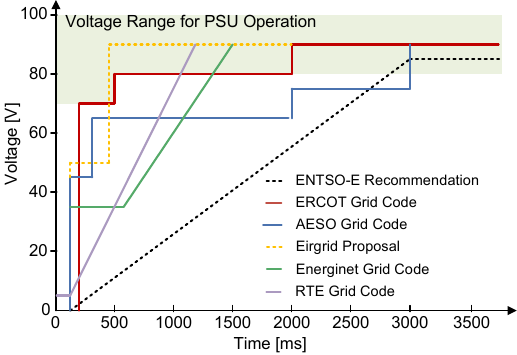}
    \caption{Planned and ongoing Fault-Ride Through (FRT) requirements for Data Centers \cite{ENTSO-E_Rec, EirGrid_FRT, AESO_FRT, Energinet_FRT, RTE_FRT}.}
    \label{fig:FRT}
\end{figure}

Fig.~\ref{fig:FRT} shows a proposal and examples of the application of the under-voltage FRT requirements in grid codes for large-scale demand, including data centers, as defined by TSOs. In December 2025, ENTSO-E issued a recommendation urging European TSOs to establish common connection requirements for data centers, including FRT requirements \cite{ENTSO-E_Rec}. As of July 2026, ERCOT, AESO, Energinet, and RTE are applying FRT requirements for loads connected via transmission lines \cite{AESO_FRT, Energinet_FRT, RTE_FRT}. EirGrid has submitted a proposal and is awaiting final approval from the regulatory authority \cite{EirGrid_FRT}. Energinet, EirGrid, and AESO require that the connection be maintained for 150 ms even if the voltage drops to 0\%. RTE requires maintaining the connection for 150 ms when the voltage drops to 5\%. RTE and Energinet relax the voltage levels at which connection maintenance is required in proportion to the elapsed time. Other grid codes establish stepped connection voltage requirements over time.

PSUs are designed to meet the connection maintenance time requirements for voltage sags established by the Information Technology Industry Council (ITIC). The ITIC requirements stipulate that operation must be guaranteed for up to 50 ms when the voltage drops to 70\% of its nominal level, and for up to 10 seconds when it drops to 80\% of its nominal level \cite{EPRI}. Therefore, data centers must implement measures to ensure stable PSU operation even during transient voltage drops that fall below these thresholds.

\subsubsection{Responsible parties for data center integration}

The responsibility of power system operators lies in ensuring transparent information disclosure and establishing new grid connection rules and technical requirements. EU has noted that insufficient transparency in disclosing information such as available grid capacity delays progress on data center projects \cite{EU_FCFS}. The EU Grid Action Plan calls for ENTSO-E and the EU DSO Entity to develop a joint portal that will collect and publish hosting capacity information across Europe, improve comparability, and support more informed investment decisions at a European scale \cite{ENTSOE2026DataCentres}.
To prevent speculative data center construction applications and the circumvention of rules, effective prioritization rules and queue management are essential in addition to capacity transparency. 

It is also necessary to update the technical requirements of grid codes for emerging loads, including data centers. In Ireland, an incident occurred on June 6, 2026, in which 480~MW, equivalent to 52\% of the country’s data center demand, was disconnected as a result of a single-phase fault at the 220~kV substation in central Dublin \cite{EirGrid_Update}. This incident shows that FRT-incompatible demand increases the risk to power-system stability.

Data center developers and operators must ensure compliance with grid codes and achieve grid congestion relief and carbon-aware data center operations. To alleviate grid congestion and enable carbon-aware operations, it is necessary to unlock the flexibility in both the physical resources and computational algorithms within the data center.

\subsection{Convergent Power-Information Infrastructures}


Going beyond the conventional paradigm, a novel insight is that the boundary between power and information infrastructures is diminishing: power and information interacts at progressive dimensions, and re-purposing either channel to carry the other's traffic enables extra flexibility for task migration within the same infrastructure, promisingly pushing the frontier of data center design compactness or sustainability. We elaborate these insights along the following directions:

\subsubsection{Cybersecurity}

Cybersecurity in data center systems spans both Information-Technology (IT) and Operational-Technology (OT) scenarios from IT racks to power supply hardware. Beyond the pure \textit{data} concerns intrinsic to the data centers themselves \cite{rawat2021bigdata} and associated energy footprints \cite{luca2017cyberenergy}, this power-information infrastructure faces operational and \textit{energy} challenges from hardware spoofing---typically false data injection attacks (FDIAs) or denial-of-service (DoS), etc \cite{sahoo2021cyberpe}---at data center infrastructure management (DCIM) platforms/facilities, the grid, as well as power electronics interfaces. Firstly, as remote coordination is increasingly deployed, exposure to the internet or similar connectivity opens the attack surface with entry points of false control commands. The consequences differ by target: attacks on data center management primarily degrade service quality or trigger operational malfunctions, whereas attacks on grids or power electronics converters propagate more rapidly at rack and system level, compromising data center operation \cite{liu2015cybergrid, huang2018cybergrid} or even pushing the system across its stability boundaries \cite{sahoo2021cyberpe, sahoo2021cyberacmg}. Secondly, considering the scale of data center loads, the impact can also run the opposite way, that abnormal behaviors induced by cyber attacks may escalate into load contingencies that threaten the resilience of local grids \cite{adrian2017cybergrid, subhash2025cybergrid}. Similarly, as smart UPS systems become more prevalent---especially when aligning the aforementioned SAFEr Grid infrastructure---their energy-management functions represent another point of concern, potentially adding to IT downtime on one side or grid-level faults on the other. It is therefore essential to understand the underlying mechanisms driving these coupled cyber-power vulnerabilities and to address them strategically at their source.

\subsubsection{Power line data transmission}

Power line data transmission, developed as power line communication (PLC) or talkative power, superimposes information onto the voltages and/or currents for power delivery ("power signals"), such that it can be detected by a receiver on another end purely through transmission lines \cite{he2020tp, marco2023tp,10909727}. This has been used, for instance, in smart metering or remote control in microgrids \cite{lopez2019plc, AFE031}. For data center specifically, where every layer of the power chain is already densely instrumented with power electronics and essentially digital controlled, power line data transfer represents a low-incremental-cost way to utilize the power layers for data transfer, defining "new" information channels that improve system compactness and add resilience against communication-network outages.

\subsubsection{Local power-over-fiber/photonic delivery}

As rack power density continues to rise, power-over-fiber addresses the resulting challenges in copper volume and electromagnetic isolation (both galvanic isolation and EMI) \cite{carneiro2024pof, arroyave2024pof}, particularly at the chip level, while photonic interconnects aim to scale data centers past the constraints of latency and bandwidth of electrical interconnects \cite{hadea2025photonic, moran2026photonic}. Both are most relevant where power and/or information density are simultaneously the first-order design requirements. Although still maturing, the two directions are worth treating as candidate solutions more than merely a convenience, serving in future data-center physics where copper-based delivery may become a binding constraint.

\subsubsection{Physical/analog computing concept}

\begin{figure}[t]
    \centering
    \includegraphics[width=0.9\linewidth]{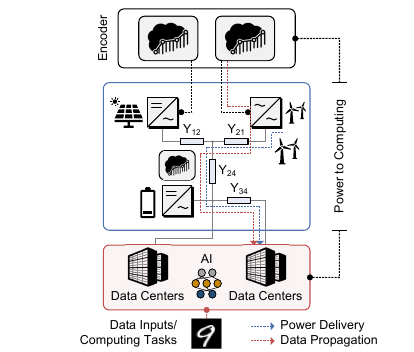}
    \caption{Conceptual illustration of computing on power infrastructure. Data centers are projected to harness the intrinsic dynamics of a power grid that perform the transformation of encoded data alongside normal power delivery, thus capitalizing on the interactivity between the utility power network and data centers.}
    \label{fig_physcompute}
\end{figure}

Another conceptually distinct yet speculative direction exploits the shared physics of power and information transformation, and exploits them algorithmically. Inspired by the physical/analog computing initiatives \cite{yan2024reservoir, wright2022pnn, momeni2025pnn}, power electronics-penetrated infrastructure can itself serve as a computing substrate---a preliminary DC-grid implementation of fundamental affine transformation is reported in \cite{phycomp}. With data encoded into system perturbations, the underlying physical laws governing the voltage-current dynamics in the network perform an inherent computing mechanism alongside normal power flow, as depicted in Fig.~\ref{fig_physcompute}. The same control logic could support outsourcing of computing load or self-referential coordination of server workload or power delivery, eventually closing the loop between the physical power layer and the computational demand it serves, with attendant efficiency and sustainability benefits.

\subsection{Sustainable Data Centers in Power Grids}

The carbon footprints of data centers have been increasingly underscored as an initiative for participating in the global energy infrastructure sustainably and responsibly. According to \cite{NBER_2026}, data center electricity consumption in the United States is projected to reach approximately 250~TWh by 2025, and monetary damages from air pollution and greenhouse gas (GHG) emissions from data centers are estimated to account for approximately 5\% of GDP. In Europe, an initiative aiming for carbon neutrality by 2030 has been formed, with more than 100 data center operators and trade associations participating. This agreement advocates for the certification of data centers' PUE, the procurement of 100\% carbon-free energy, water conservation, and the reuse of waste heat \cite{CNDCP}.

In this section, we introduce ongoing efforts and outlooks on reducing the carbon footprints of data centers, including carbon-aware computing and sustainability initiatives for data center equipments. An overview of the cost allocation for these decarbonization measures is also provided.


\subsubsection{Carbon footprint of computing tasks and carbon-aware computing}

GHG emissions from data center operations are determined by the GHG emission intensity of the electricity used to power the data centers, which is a factor that varies by time and location. Google actually allocates workloads to minimize GHG emissions from electricity used by its data centers through the Carbon
Intelligent Computing System (CICS), designed to minimize GHG emissions from the electricity used by data centers. The system predicts data center demand and the grid’s carbon intensity one day in advance and aims to reduce CO\textsubscript{2} emissions by flexibly allocating workloads across the data center fleet. In this example, enabling CICS has been shown to reduce average cluster power consumption by 1–2\% during the highest-carbon hour \cite{Radovanovic_2022}.
Research in \cite{Hall_2025} proposes methods to implement carbon-aware computing for tasks such as offline data processing, model training, and simulation pipelines.

Research in \cite{Oliveira_2026} evaluates the carbon footprint across the AI lifecycle. According to this research, the proportion of emissions attributable to computing tasks and hardware varies depending on model size. In particular, for large-scale models, the study shows that operational inference has a higher carbon footprint than training. Research in \cite{Han_2026} proposes an algorithm for carbon-aware scheduling of training and inference workloads across multiple AI data centers. Simulation confirms that scheduling across three AI data center parks, while considering continuity-constrained workloads, can decrease operating costs by at least 1.69\% and carbon emissions by 4.5\% compared to four alternative schemes, while maintaining computational tractability.

\begin{figure}[t]
    \centering
    \includegraphics[width=\linewidth]{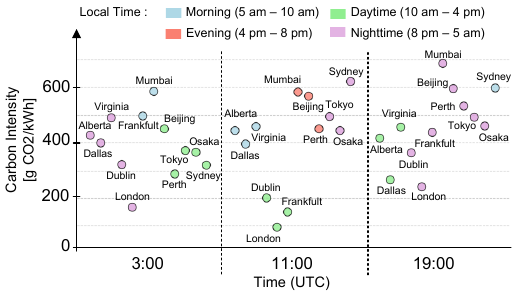}
    \caption{Comparison of carbon intensity of electricity in major data center hub cities by Coordinated Universal Time (UTC), August 12th, 2026, obtained from Electricity Maps \cite{ElectricityMap}.}
    \label{fig:carbon}
\end{figure}

Furthermore, it is important to consider the allocation of computational tasks while accounting for two constraints: grid congestion and carbon footprint. Fig.~\ref{fig:carbon} compares the carbon intensity of major data center hub cities, listed in Table~\ref{tab:spatial}, by Coordinated Universal Time (UTC), using data from Electricity Maps \cite{ElectricityMap}. Three time slots, 3~AM, 11~AM, and 7~PM on August 12th, 2026, are selected. During the daytime, when solar power generation increases, carbon intensity decreases compared to the evening and nighttime. Therefore, at night, when carbon intensity is higher, allocating workloads to regions where it is daytime at the same time is an effective way to reduce emissions associated with computing.

\subsubsection{Cost allocation for carbon-aware computing}

Cost allocation among stakeholders in carbon-aware computing varies by data center type. In data centers owned by cloud service providers, such as hyperscale data centers, the service provider can be responsible for managing the data center’s power consumption and carbon footprint. In fact, Google has demonstrated carbon-aware computing across its fleet of owned data centers \cite{Google_2023}. Additionally, cloud service providers can include carbon footprint-related charges in their cloud service fees. Even in enterprise data centers built by companies for their own use, the companies themselves can manage the data center’s carbon footprint.
On the other hand, in the case of colocation data centers, the data center operator enters into a power supply contract, and that cost is incorporated into the usage fees charged to client companies. Methods for allocating the costs of carbon-aware computing among multiple client companies using computing resources have been proposed, such as data center operators offering voluntary bidding programs to tenants \cite{Islam_2012}.

\section{Conclusion}

AI data centers are emerging as large, dynamic, and power electronics-dominated electrical infrastructures whose scalable deployment cannot be assessed through energy consumption and peak demand alone. Their grid impact is determined by the coupled behavior of computing workloads, power-delivery architectures, cooling systems, and energy buffers. This article provides a comprehensive overview of the resulting challenges and trends from grid connection to chip-level regulation. It first shows that interconnection capacity, equipment lead times, and evolving grid codes increasingly constrain data-center deployment, while workload orchestration, cooling, on-site resources, and energy buffers can provide conditional spatio-temporal flexibility. It further highlights the transition toward higher-voltage DC architectures, solid-state transformers, wide-bandgap devices, advanced chip-level power delivery, and liquid cooling. The proposed rack--facility--system framework clarifies that stability must be addressed across multiple boundaries: workload-induced variations and constant power load effects can challenge rack-level DC buses; converter interactions and mode transitions can affect facility-level operation; whereas fault ride-through, load transfer, disconnection, and restoration can affect grid voltage and frequency.

Several future trends are outlined in this article. For data center operators, compute scheduling, energy buffering, converter control, and reliability management should be coordinated to define a predictable power envelope at the point of common coupling, including ramping, flexibility, fault ride-through, recovery behaviors, cyber-physical resilience, and carbon awareness. For system operators and regulators, AI data centers should be represented as dynamic, potentially flexible resources rather than fixed demand, with connection arrangements and grid codes that recognize verifiable flexibility, power quality, fault ride-through, and staged post-fault restoration. 
Ultimately, grid-to-chip co-design provides a principled pathway for scaling AI infrastructure from grid integration challenges into a controllable, reliable and efficient \textit{grid-supportive resource}, from which the developments of AI data centers and power grids are progressively aligned.




\bibliography{references.bib}
\bibliographystyle{IEEEtran}

\balance


 





\end{document}